\documentclass[a4paper,12pt,reqno]{amsart}

\usepackage{amsmath}
\usepackage{amssymb}
\usepackage{amsfonts}
\usepackage[all,2cell]{xy} \UseAllTwocells \SilentMatrices
\usepackage{geometry,latexsym,amsmath,amstext,mathrsfs}
\usepackage{amssymb,amsthm,amscd,amsopn,verbatim,graphics}
\usepackage[english]{babel}

\theoremstyle{plain}
\newtheorem{theorem}{Theorem}
\newtheorem{lemma}[theorem]{Lemma}
\newtheorem{proposition}[theorem]{Proposition}
\newtheorem{corollary}[theorem]{Corollary}
\theoremstyle{definition}
\newtheorem{definition}[theorem]{Definition}
\theoremstyle{remark}
\newtheorem{example}[theorem]{Example}
\newtheorem{remark}[theorem]{Remark}

\numberwithin{theorem}{section}
\numberwithin{equation}{section}

\def\P2{{P^{[2]}}}

\def\11{{\mbox{\boldmath $1$}}}

\def\eq{\begin{equation}}
\def\en{\end{equation}}

\def\eq#1\en{\begin{equation}#1\end{equation}}
\def\eqa#1\ena{\begin{eqnarray}#1\end{eqnarray}}

\begin{document}

\baselineskip=1.2\baselineskip

\title[AKSZ construction of topological open p-brane action ...]{AKSZ
construction of topological open p-brane action and Nambu brackets}

\author[P Bouwknegt]{Peter Bouwknegt}
\address[P Bouwknegt]{
Department of Theoretical Physics,
Research School of Physics and Engineering, and
Department of Mathematics, Mathematical Sciences Institute,
The Australian National University,
Canberra, ACT 0200, Australia}
\email{peter.bouwknegt@anu.edu.au}

\author[B Jur\v co]{Branislav Jur\v co}
\address[B Jur\v co]{
Mathematical Institute, Charles university,
Sokolovsk\'a 83, Prague 186~75,
Czech Republic}
\email{jurco@karlin.mff.cuni.cz, branislav.jurco@googlemail.com}

\begin{abstract}
We review the AKSZ construction as applied to the topological open
membranes and Poisson sigma models. We describe a generalization to
open topological $p$-branes and Nambu-Poisson sigma models.
\end{abstract}


\maketitle


\section{Introduction}
The purpose of this paper is two-fold. First, we review the
Alexandrov-Kontsevich-Schwarz-Zaboronsky (AKSZ)
formulation of topological open membranes and Poisson sigma models,
following \cite{Park}, \cite{HofmanPark} and
\cite{CattaneoFelderAKSZ}. Second, we propose a generalization to
the case of topological open $p$-branes and Nambu-Poisson sigma
models.

The Poisson sigma model, introduced in \cite{StroblSchaller} and
\cite{Ikeda}, and its twisted version introduced in
\cite{KlimcikStrobl}, play an important role in modern mathematical
physics. The most striking application of the open Poisson sigma
model is the path integral derivation of Kontsevich's formality, in
particular of his celebrated star product \cite{Ko}, in \cite{CF}.
For this, it was necessary to use BV formalism. Within this
formalism Kontsevich's formality appears as a consequence of the
Ward identities for the BV quantized open Poisson sigma model.
Another example of the use of Poisson sigma models is the
integration of Poisson manifolds to symplectic groupoids in
\cite{CFSGroupoid}.

The AKSZ formalism \cite{AKSZ} is a geometric formalization of the
BV formalism. It leads to a powerful method for constructing BV
actions starting from geometric data, the super worldvolume and the
superspace. The super worldvolume is a differential graded manifold,
equipped with a measure invariant under the cohomological vector
field and the super spacetime a differential graded symplectic
manifold, for which the cohomological vector filed is a Hamiltonian
one. The AKSZ formalism is a prescription how to construct, starting
with these data, a solution to master equation on the space of the
corresponding superfields. Examples of the construction comprise the
BF model in any dimension, the 2 dimensional A- and B-model, 3
dimensional Chern-Simons \cite{AKSZ}, \cite{Park} and 3 dimensional
Rozansky-Witten theory \cite{QZ}. Further applications of the AKSZ
construction can be found, e.g., in \cite{Ikeda3}, \cite{Ikeda4} and
\cite{FS}. The AKSZ formulation of the Poisson sigma model was given
in \cite{CattaneoFelderAKSZ}. Papers \cite{Park} and
\cite{HofmanPark} describe an AKSZ formulation of open topological
membranes in the presence of a closed 3-form and a twisted Poisson
tensor, which leads on shell to the twisted Poisson sigma model. The
BV quantization of such topological membranes is described in
\cite{HofmanPark1}. In \cite{HofmanPark} the authors also describe
the relation of topological open membranes to Lie and Courant
algebroids and Dirac structures, cf. also \cite{Roytenberg},
\cite{Ikeda2}, for some related results about 3 dimensional Courant
sigma model.

Nambu-Poisson structures are the most natural generalizations of
Poisson structures.  The original definition of the Nambu bracket of
order 3 goes back to the Nambu's seminal paper \cite{Nambu}. The
generalization and the modern geometric formulation is due to
Takhtajan \cite{Takhtajan}. An order $p$ Nambu-Poisson bracket on a
manifold $M$ determines  the structure of a Filippov $p$-algebra  on
$C^\infty(M)$ \cite{Fil}. The recent interest in such generalized
structures was motivated by the Bagger-Lambert-Gustavsson model for
M2-branes \cite{BH}, \cite{BL1}, \cite{BL2}, \cite{Gus}. More
recently, the relevance of Nambu-Poisson brackets of order 3 for the
description of M5-branes was noticed in \cite{Ho1}.

{}From the above discussion, it seems to be natural (and hopefully
useful) to search for a proper generalization of the above mentioned
works on topological open membranes and Poisson sigma models to the
case of topological open $p$-branes and Nambu-Poisson sigma models.
We also hope that the topological open $p$-branes can be quantized
using the BV quantization scheme, similarly to topological open
membranes \cite{HofmanPark1}. This could possibly lead to a kind of
deformation quantization of Nambu-Poisson structures\footnote{For
deformation quantization of Nambu-Poisson structures see
\cite{Dito}. Interested reader can consult, e.g., the review
\cite{Azcarraga} for a short discussion of other approaches to the
quantization of Nambu-Poisson structures.}.

This paper is organized as follows. In Section 2 we summarize, for
the reader's convenience, the relevant facts regarding the Dorfman
(Courant) bracket on sections of $TM \oplus \wedge^{p-1}T^*M$.  The
discussion includes the twistings of a $p$ Dorfman bracket by a
$p+1$-form and a $p$-vector field, which leads to a generalization
of the so called Roytenberg bracket. Most of the material of this
section can be found in \cite{Hagiwara} and \cite{Sheng}.

In Section 3, we recall the basic facts concerning Nambu-Poisson
structures. In particular, we describe different, but equivalent,
forms of the fundamental identity (FI) and recall the relation
between the Dorfman and Poisson-Nambu brackets following, e.g.\
\cite{Hagiwara} and \cite{Sheng}. We discuss the possibility of
twisting a Nambu-Poisson bracket of order $p$ with a closed $p+1$
form, which, as it turns out, is not possible for $p\geq3$ (this is
probably known to the experts, but we could not find it anywhere the
the existing literature). Further, we describe gauge equivalence of
Nambu-Poisson structures and relate it to the higher (semiclassical)
version of the Seiberg-Witten map, well-known from non-commutative
gauge theory. Our discussion of the Seiberg-Witten map is a
generalization of the low dimensional cases $p=2$ in \cite{Jurco1,
Jurco2}, and $p=3$ in \cite{Ho}.

In Section 4, we collect the relevant material about differential
graded (dg) symplectic manifolds, which are the natural framework
for the Batalin-Vilkovisky formalism. Among various relevant
examples, we describe twisted Dorfman brackets within the framework
of dg symplectic manifolds. Here we closely follow the review
\cite{CattaneoSch}, which is based on \cite{Roytenberg}.

The AKSZ construction of BV actions is reviewed in Section 5
following, with small modifications, \cite{CattaneoSch} and
\cite{HofmanPark}. As already mentioned above, the AKSZ formalism
\cite{AKSZ} provides for a solution of the classical master equation
 on the space of maps between a super worldvolume and
a super spacetime under rather mild assumptions. Since we are
interested in open $p$-branes, we describe in detail the effect of
canonical transformations on the boundary conditions and the
boundary terms too. This is an important point in understanding the
construction of actions described in this paper.

In Section 6, we provide examples of the AKSZ construction. The
first one is the Poisson sigma model, following
\cite{CattaneoFelderAKSZ}, the second one is the open topological
membrane and the twisted Poisson sigma model, following \cite{Park}
and \cite{HofmanPark}. The open topological membrane action has a
WZW-type bulk term originating from a closed $3$-form $c$ and a
boundary term originating from a Poisson tensor $\pi$ twisted by
$c$. There is a gauge symmetry which changes the $3$-form $c$ by an
exact piece and the twisted Poisson structure to an equivalent one,
hence relating the bulk and boundary interactions. This gauge
symmetry is closely related to the semiclassical Seiberg-Witten map.
The twisted Poisson sigma model is obtained from the open
topological membrane on shell.

The open topological membrane and the twisted Poisson model are
generalized to an open topological $p$-brane and a Nambu-Poisson
sigma model (recall, there the twisting of a Nambu-Poisson structure
doesn't work for $p>2$) in Section 7. Again, the construction is
based on AKSZ. However, the generalization is not a straightforward
one. For instance, it does include the case $p=2$ in a non-trivial
way. Namely, for $p=2$, we have twice as many fields compared to the
action of the topological open membrane and the latter is obtained
only after imposing further constrains on the fields. Nevertheless,
also for general $p$, the resulting topological open $p$-brane
action contains a $p+1$-form $c$ coupled to the bulk through the WZW
term and a Nambu-Poisson tensor coupled to the boundary. Also here,
we have a gauge symmetry which changes the $p+1$-form $c$ by an
exact piece and Nambu-Poisson structure to a gauge equivalent one
(in the sense of Section \ref{NPstructures}), hence relating the
bulk and boundary interactions. This gauge symmetry is closely
related to the higher semiclassical Seiberg-Witten map as described
in Section \ref{NPstructures}. On shell we have a generalization of
the Poisson sigma model, which we call the Nambu-Poisson sigma
model. This model is also obtained quite naturally as the
topological limit of models arising in the study of p-brane actions
with background fields \cite{JS}. Further generalizations,
properties and applications of the Nambu-Poisson sigma model will be
discussed in this forthcoming paper.

\section{Dorfman brackets}\label{SectionDorfman}

Let $M$ be a smooth finite-dimensional manifold,
and let $E$ denote a vector bundle over $M$. The set of sections of $E$
will be denoted by $\Gamma E$.  In particular we have the tangent bundle $TM$,
whose sections $\mathfrak X(M) = \Gamma(TM)$ are vector fields, and the cotangent
bundle $T^*M$ whose sections $\Gamma(T^*M)$ are 1-forms.
We also denote by $\mathfrak X^p(M) = \Gamma(\wedge^p TM)$, and
$\Omega^p(M) = \Gamma(\wedge^p T^*M)$ the set of $p$-vector fields
and $p$-forms, respectively.

In this section we collect some basic facts concerning the Dorfman bracket
on $\Gamma(TM\oplus \wedge^{p-1}T^{\ast}M) = \mathfrak X(M) \oplus \Omega^{p-1}(M)$.
We have
the following definition \cite{Hagiwara} (see also \cite{Hitchin},
\cite{Gualtieri}, \cite{Sheng}, \cite{Marsden}).\footnote{Let us
note, that almost everything in this and the subsequent sections can
be formulated more generally, by replacing the tangent bundle $TM$
and cotangent bundle $T^*M$ by a Lie algebroid $A$ over $M$ and its
dual $A^*$, respectively \cite{Wade}.} Everything in this section,
with an exception of maybe Remark \ref{1.4} and Proposition
\ref{1.2}, can be found, e.g., in \cite{Sheng}.

\begin{definition}
The \emph{Dorfman bracket} of order $p\geq 2$ on
sections of $E= TM\oplus \wedge^{p-1}T^{\ast}M$ is defined by
$$
[(X, \alpha), (Y, \beta))] = ([X, Y],\mathcal{L}_X\beta - \iota_Yd\alpha) \,,
$$
where $X, Y \in \mathfrak X(M)$, and $\alpha, \beta\in \Omega^{p-1}(M)$.
\end{definition}

Let $\rho$ denote the projection $\rho: E\to TM$, the so-called
{\it anchor map},  and let
$\langle\cdot ,\cdot\rangle$ denote the
$\Omega^{p-2}(M)$-valued non-degenerate bilinear pairing between $\mathfrak X(M)$ and
$\Omega^{p-1}(M)$ given by
$$
\langle(X,\alpha), (Y,\beta)\rangle = \frac{1}{2}(\iota_X\beta +
\iota_Y\alpha) \,.
$$
The important properties of the Dorfman bracket $[\cdot,\cdot ]$
on $\Gamma E = \Gamma(TM\oplus \wedge^{p-1}T^{\ast}M)$ are
summarized in the following theorem\footnote{We will use the same
notation, $\rho$, for the mapping induced on the sections, $\rho:
\Gamma(E)\to \Gamma(TM)$.}

\begin{theorem} For $e_1, \, e_2, \, e_3\in \Gamma E$ and
$f\in C^\infty (M)$
\begin{equation} \label{eqnBa}
\rho[e_1, e_2]= [\rho e_1, \rho e_2] \,,
\end{equation}
\begin{equation} \label{eqnBb}
[e_1, fe_2]= f[e_1, e_2]+ (\rho(e_1)f)e_2 \,,
\end{equation}
and the bracket is a Leibniz bracket, i.e.\ the Jacobi identity
\begin{equation} \label{eqnBc}
[e_1,[e_2, e_3]= [[e_1, e_2], e_3]+[e_2, [ e_1,e_3]] \,,
\end{equation}
is satisfied.
The pairing and the bracket are compatible, i.e.,
\begin{equation} \label{eqnBd}
\mathcal{L}_{\rho(e_1)}\langle e_2, e_3 \rangle = \langle [e_1, e_2],
e_3\rangle + \langle e_2, [e_1,e_3] \rangle \,.
\end{equation}
\end{theorem}

Let $c$ be a $(p+1)$-form on $M$. Then the Dorfman bracket $[\cdot,\cdot]$
can be twisted by $c$.
\begin{definition}
The twisted Dorfman bracket is defined as $$[(X, \alpha), (Y,
\beta))]_c := ([X, Y],\mathcal{L}_X\beta - \iota_Yd\alpha + \iota_{X\wedge
Y}c) \,.
$$
\end{definition}
We have the following proposition
\begin{proposition} The twisted Dorfman
bracket is a Leibniz bracket iff $c$ is closed; $dc=0$. Further, for
any $b \in \Omega^{p}(M)$ define $e^b: E\to E $ as
$$
e^{b}(X,\alpha) := (X, \alpha + \iota_Xb) \,.
$$
We have
$$
e^b[e_1, e_2]_{c + db}= [e^b e_1, e^b e_2]_c \,.
$$
\end{proposition}

\begin{remark}
The Dorfman bracket (twisted or not) is not antisymmetric; its
antisymmetrization is called the (twisted) \emph{Courant bracket}.
However, the  Courant bracket (twisted or not) does not obey the
Jacobi identity. The triple $(E, [\cdot,\cdot]_{c}, \rho)$ has the
structure of a Leibniz algebroid (properties \eqref{eqnBa}-\eqref{eqnBc})
\cite{Hagiwara}, \cite{Sheng}.
Moreover, taking into account the differential $d:
\Omega^{p-2}(M)\to \Omega^{p-1}(M)\oplus \mathfrak X(M)$, defined by $d:\alpha \mapsto
(d\alpha, 0)$, the quadruple $(E, [\cdot,\cdot]_{c}, \langle \cdot,\cdot \rangle, d)$ has
a structure of a (Courant-) Dorfman algebra \cite{Roytenberg3}, \cite{Ekstrand}.
\end{remark}

\begin{remark} \label{1.4}
The Dorfman bracket can also be twisted by a
$p$ multi-vector field $\zeta \in \mathfrak X^p(M)$.
Define $e^{\zeta}: E\to E $ as\footnote{Henceforth, for any $\zeta\in\mathfrak X^p(M)$, and
$b\in \Omega^p(M)$, we define maps $\zeta^\sharp: \Omega^{p-1}(M) \to \mathfrak X(M)$,
and $b_\sharp: \mathfrak X(M) \to \Omega^{p-1}(M)$ by
$\zeta^\sharp (\alpha) = \imath_\alpha \zeta$, and $b_\sharp(X) = \imath_X b$,
for all $\alpha\in\Omega^{p-1}(M)$, and $X\in\mathfrak X(M)$, respectively.}
$$
e^{\zeta}(X,\alpha) := (X + \zeta^\sharp\alpha, \alpha) \,,
$$
and introduce the $(c,\zeta)$-twisted Dorfman bracket
$$
e^\zeta[e_1, e_2]_{(c,\zeta)}= [e^\zeta e_1, e^\zeta e_2]_c \,.
$$
In the case of $p=2$, it is due to Roytenberg \cite{RoytenbergBracket}.
\end{remark}

\begin{proposition} \label{1.2} Assume that the map $(1 + b_\sharp \circ \zeta^\sharp):
\Omega^{p-1}(M) \to \Omega^{p-1}(M)$ is
invertible and denote $\zeta'{}^\sharp= \zeta^\sharp \circ (1+ b_\sharp
\circ\zeta^\sharp)^{-1}$. We have
$$
e^b e^\zeta = e^{\zeta'}e^{(b,\zeta)} \,,
$$
where $e^{(b,\zeta)}(X,\alpha) = ((1 - \zeta'{}^\sharp\circ b_\sharp)(X),
(1 + b_\sharp \circ \zeta^\sharp)(\alpha) + b_\sharp (X))$.
\end{proposition}

\section{Nambu-Poisson structures}\label{NPstructures}

The original definition of the Nambu bracket of order 3 goes back to the
Nambu's seminal paper \cite{Nambu}. The  generalization and the
modern geometric formulation is due to Takhtajan \cite{Takhtajan}
(for a review on $n$-ary algebras, see \cite{Azcarraga}).

\begin{definition} A Nambu-Poisson bracket of order $p\geq 2$
on $M$ is a skew-symmetric $p$-linear map $\{\cdot ,\ldots,\cdot\}: C^\infty(M)
\times\ldots\times C^\infty (M)\to C^\infty(M)$ having the Leibniz
property
$$
\{f g,f_2 \ldots,f_{p}\}=f\, \{g,f_2, \ldots,f_{p}\} + g\, \{f,f_2, \ldots,f_{p}\} \,,
$$
and satisfying the so called fundamental identity (FI)
\begin{equation}\label{FI1}
\{f_1, \ldots,f_{p-1},\{g_1,  \ldots,g_{p}\}\} =
\sum_{i=1}^{p}\{g_1,\ldots \{f_1,\ldots,g_{i}, \ldots,f_{p-1}\},
\ldots,g_{p}\} \,.
\end{equation}
\end{definition}
In terms of the corresponding $p$-vector field $\pi\in\mathfrak X^p(M)$, defined by
$$\pi(df_1\wedge\ldots\wedge df_p\} :=\{f_1, \ldots,f_{p}\}\,,$$ the
fundamental identity can be expressed as in the following
proposition, which  can be
read off from,  e.g., \cite{Hagiwara}, \cite{Sheng} (cf.\ also
\cite{Ibanez}).

\begin{proposition}
A Nambu-Poisson bracket of order $p$ on $M$ is uniquely determined
by a multi-vector field $\pi$ of order $p$, satisfying either of the
following:
\begin{itemize}
\item[(i)] for all $\alpha, \beta\in \Omega^{p-1}(M)$,\footnote{Equivalently, as an equality
of maps from  $\Omega^{p-1}(M)$ to $\mathfrak X(M)$ in the form
$[\pi^\sharp\alpha, \pi]_S{}^\sharp  = (\mathcal{L}_{\pi^\sharp \alpha}\pi)^\sharp = -\pi^\sharp\circ (d\alpha)_\sharp\circ
\pi^\sharp$. \label{FIfootnote}}
$$ [\pi^\sharp\alpha, \pi ]_S{}^\sharp \beta= (\mathcal{L}_{\pi^\sharp \alpha}\pi)^\sharp \beta = -
\pi^\sharp(\iota_{\pi^\sharp\beta}d\alpha) \,.
$$
where $[\cdot,\cdot]_S$ denotes the Schouten
bracket of multi-vector fields.
\item[(ii)] for $f_1,\ldots, f_{p-1} \in  C^\infty(M)$,
$$ [\pi^\sharp (df_1\wedge
\ldots \wedge df_{p-1}), \pi]_S = \mathcal{L}_{\pi^\sharp (df_1\wedge\ldots\wedge
df_{p-1})}\pi =0 \,,
$$
\item[(iii)] for $(p-1)$-forms $\alpha$ and $\beta$,
\begin{equation}\label{integrab}
[\pi^\sharp \alpha, \pi^\sharp \beta] = \pi^\sharp [\alpha,\beta]_\pi \,,
\end{equation}
where $$ [\alpha , \beta]_\pi  := \mathcal{L}_{\pi^\sharp \alpha}\beta  - \iota_{\pi^\sharp\beta}
d\alpha \,.
$$
\end{itemize}
\end{proposition}

\begin{corollary} In some local coordinates $x^i$ on $M$, the
fundamental identity gives two conditions, an algebraic and a
differential one. The algebraic one is
\begin{equation}\label{alg1}
\Sigma_{i_1\ldots i_p j_1\ldots j_p} = \Sigma_{j_1 i_2\ldots i_p i_1
j_2\ldots j_p} \,,
\end{equation}
where
\begin{equation*}\Sigma_{i_1\ldots i_p j_1\ldots
j_p}=\frac{1}{p!}\epsilon_{l_1\ldots l_{p+1}}^{i_p j_1\ldots
j_p}\pi^{i_1\ldots i_{p-1}l_1}\pi^{l_2\ldots l_{p+1}} \,,
\end{equation*}
and the differential one is
\begin{equation}\label{alg2}
\pi^{i_1\ldots i_{p-1}k} \partial_k\pi^{j_1\ldots
j_{p}}=\frac{1}{(p-1)!}\epsilon_{l_1\ldots l_{p}}^{j_1\ldots
j_p}\pi^{ l_1\ldots l_{p-1}k}\partial_k(\pi^{i_1\ldots i_{p-1}l_p}) \,.
\end{equation}
\end{corollary}

\begin{remark}\label{decompRemark}
The algebraic condition (\ref{alg1})  assures that the second order
derivative terms in the fundamental identity (\ref{FI1}) vanish,
which is a nontrivial statement for $p\geq 3$. This condition is
equivalent, for $p\geq 3$, to the \emph{decomposability} of the
Nambu-Poisson tensor \cite{Gautheron}, \cite{Alekseevsky}, (cf. also
\cite{Panov}, \cite{Marmo}, \cite{Nakanishi})\footnote{In
\cite{Gautheron}, an observation that the decomposability follows
from the so called Weitzenb\"ock condition is attributed to L.
Takhtajan.}. More precisely: Let us fix a point $x\in M$ for which
$\pi(x)\neq 0$, then, locally around $x$, $\pi=v_1\wedge \ldots
\wedge v_p$, with some local vector fields $v_1,\ldots, v_p$. Let us
also note that any $p$-vector field of such a form trivially
fulfills the fundamental identity. If $V_x\subset T_x M$ denotes
the $p$-dimensional subspace generated at the point $x\in M$ by
($v_1,\ldots, v_p)$, then the fundamental identity in the form
(\ref{integrab}) assures that the field of subspaces $V_x$ is
integrable.
\end{remark}

\begin{corollary}
Due to the decomposability, for $p\geq 3$, the fundamental identity
in the form of footnote \ref{FIfootnote} can be rewritten as
$$
[\pi^\sharp \alpha, \pi]_S{}^\sharp = - \pi^\sharp \circ (d\alpha)_\sharp \circ  \pi^\sharp
= (-1)^p\,  (\imath_{d\alpha} \pi)  \pi^\sharp \,.
$$
\end{corollary}

The following characterization of a decomposability of a $p$-vector,
$p\geq 3$, will be useful later\footnote{See \cite{Schouten}, where
also the equivalence to the Weitzenb\"ock condition is shown.}.
\begin{lemma}\label{decompLemma}A $p$-vector $\pi$ is decomposable iff
\begin{equation}
\pi^{[i_1\ldots i_p}\pi^{j_1]\ldots j_p}=0 \,,
\end{equation}
where the square brackets denote antisymmetrization.
\end{lemma}

The relation between the Dorfman bracket on $\Gamma(TM\oplus \wedge^{p-1}T^*M)$
and Nambu-Poisson structures is as
follows, see, e.g., \cite{Hagiwara}, \cite{Sheng}.
\begin{theorem}
Let $\pi$ be a $p$-vector field. Its graph, {\rm graph}$(\pi)=
\{(\pi^\sharp\alpha,\alpha),\ \alpha \in \Omega^{p-1}(M)\} \subset
\mathfrak X(M) \oplus\Omega^{p-1}(M)$, is closed under the Dorfman bracket
iff $\pi$ is a Nambu-Poisson vector of order $p$.
\end{theorem}

\begin{remark}
Let $c$ be a closed $(p+1)$-form. One can try to introduce a Nambu-Poisson
structure twisted by $c$,  in analogy with the
twisted Poisson bracket for $p=2$ \cite{Severa}.
\end{remark}

Again, let $\pi$ be a $p$-vector field and let us determine when its graph, ${\rm
graph}(\pi)\subset \mathfrak X(M)\oplus\Omega^{p-1}(M)$, is closed under the
twisted Dorfman bracket. We will find, similarly to
(\ref{integrab}), the following (necessary and sufficient)
condition. For $(p-1)$-forms $\alpha$ and $\beta$,
$$
[\pi^\sharp\alpha, \pi^ \sharp \beta] = \pi^\sharp [\alpha,\beta]_{\pi,c} \,,\label{FI1Tw}
$$
where now
$$[\alpha , \beta]_{\pi,c} := \mathcal{L}_{\pi^\sharp\alpha}\beta  -
\iota_{\pi^\sharp\beta} d\alpha + \iota_{\pi^\sharp\alpha\wedge
\pi^\sharp\beta}c  \,. \label{BracketFormsTw}
$$
Equivalently,  we have for $f_1, \ldots,f_{p-1}, g_1,
\ldots,g_{p}\in C^{\infty}(M)$
\begin{align}
\{f_1, \ldots,f_{p-1},\{g_1,  \ldots,g_{p}\}\}&=
\sum_{i=1}^{p}\{g_1,\ldots, \{f_1,\ldots,g_{i}, \ldots,f_{p-1}\},
\ldots,g_{p}\} \nonumber\\ &+ c(X_{f_1,\ldots, f_{p-1}}\wedge X_{g_1} \wedge \ldots \wedge
X_{g_{p-1}}\wedge X_{g_p}) \,,
\end{align}
where $X_{f_1,\ldots, f_{p-1}} \in \mathfrak X(M)$ denotes the Hamiltonian vector field associated
to functions $f_1,\ldots, f_{p-1}\in C^\infty(M)$, defined by
$$X_{f_1,\ldots,
f_{p-1}}(h)=X_h(f_1,\ldots, f_{p-1}):=\{f_1, \ldots f_{p-1}, h\} \,.
$$

Now, following Remark \ref{decompRemark}: In some local
coordinates $x^i$ on $M$, the twisted fundamental identity again
gives two conditions, an algebraic and a differential one. The
algebraic one is identical to (\ref{alg1}) and the differential one
(\ref{alg2}) gets a contribution
coming from the closed $p+1$-form $c$, which is proportional to\\
$c_{l_1\ldots l_{p+1}}\pi^{i_1\ldots i_{p-1}l_1}\pi^{j_1\ldots
j_{p-1}l_2}\pi^{l_3\ldots l_{p+1}j_p}$.

Since, for $p\geq3$, the algebraic condition is the same as in the untwisted case,
it is again equivalent to the \emph{decomposability}
of the tensor $\pi$. Let us also note that for any $p$-vector field
of such a form, the above mentioned contribution to the differential
condition $c_{l_1\ldots l_{p+1}}\pi^{i_1\ldots
i_{p-1}l_1}\pi^{j_1\ldots j_{p-1}l_2}\pi^{l_3\ldots l_{p+1}j_p}$
vanishes identically. Hence, $\pi$ also fulfills the untwisted
fundamental identity. We can conclude that for a $p$-vector field,
$p\geq 3$, from the twisted Dorfman bracket we only get an
``ordinary'' Nambu-Poisson tensor. For $p=2$, however, we get a
twisted Poisson tensor. We have the following corollary.

\begin{corollary}\label{EquivCoro}
Let $\pi$ be a $p$-tensor and $b$ a $p$-form. Then $e^{b}({\rm graph}(\pi))$
corresponds to a graph of a $p$-vector $\pi^b$ iff $(1 +
b_\sharp\circ\pi^\sharp)$ is invertible on $\Omega^{p-1}(M)$. On
$\Omega^{p-1}(M)$, we have
$$\pi^b{}^\sharp=\pi^\sharp\circ(1 + b_\sharp\circ\pi^\sharp)^{-1} \,.$$
For $p= 2$, if $\pi$ is a Poisson bracket twisted by $c$, then
$\pi^b$ is a Poisson tensor twisted by $c - db$. If $\pi$ is a
Nambu-Poisson tensor, for $p\geq 3$,  due to the decomposability, we
have
$$
\pi^b =(1 +(-1)^{p-1}b(\pi))^{-1}\pi \,,
$$
in which case $\pi^b$ is again a Nambu-Poisson
tensor.\footnote{Recall that, due to the decomposability,
multiplying a Nambu-Poisson tensor by a smooth function gives again
a Nambu-Poisson tensor.} We say that $\pi^b$ and $\pi$ are gauge
equivalent.
\end{corollary}

\begin{remark}\label{diffequation}
Let us note that, if it makes sense, $\pi_t^b{}^\sharp :=\pi^\sharp\circ(1 + tb_\sharp\circ\pi^\sharp)^{-1}$, with
$t\in [0,1]$ is a solution to the differential equation
$$\dot{\pi}_t^\sharp =-\pi_t^\sharp \circ b_\sharp \circ\pi_t^\sharp\,,$$ interpolating between $\pi$ and $\pi^b$.
\end{remark}

\subsubsection*{Seiberg-Witten map}

In case of an exact $b$, $b=da$, we have the so called
Seiberg-Witten map (see, e.g., \cite{Jurco1}, \cite{Jurco2} for the
case of a Poisson structure and \cite{Ho} for the case $p=3$). The
Seiberg-Witten map is a (formal) diffeomorphism relating the
Nambu-Poisson tensors $\pi$ and $\pi^{da}$. We have the following
general proposition and its corollary, due to the decomposability,
valid for $p\geq 3$.

\begin{proposition}\label{SW}
Suppose that two Nambu-Poisson $p$-tensors $\pi$ and $\pi^{da}$ on
$M$ are gauge equivalent, the gauge equivalence being given by an
exact $p$-form $b=da$, such that, for $t\in[0,1]$, the map $(1 + tb_\sharp
\circ\pi^\sharp)$ is invertible on $\Omega^{p-1}(M)$ and the vector field
$\pi_t^\sharp a$, where $\pi_t^\sharp$ is defined by $\dot{\pi}_t^\sharp=
-\pi_t^\sharp\circ b_\sharp\circ\pi_t^\sharp,\,\,\pi_0^\sharp=\pi^\sharp$, is complete. Then there
exists a Nambu-Poisson map relating $\pi$ and $\pi^{da}$.
\end{proposition}

\noindent\emph{Proof.} Let us consider a one-parameter family of $p$-tensors
$\pi_t$ defined by
$$\pi_t^\sharp=\pi^\sharp\circ(1 + tb_\sharp\circ\pi^\sharp)^{-1}\,.$$
We have $\pi_0=\pi$ and $\pi_1=\pi^{da}$. Moreover, it is
straightforward to check that
$$\dot{\pi}_t^\sharp=-\pi_t^\sharp\circ b_\sharp\circ\pi_t^\sharp= (\mathcal{L}_{\pi_t^\sharp a}\pi_t)^\sharp\,.$$
Denote the corresponding flow by $\phi_t$. The map $\phi:= \phi_1$
is the sought Nambu-Poisson map. \qed

\begin{corollary}
Suppose that two Nambu-Poisson $p$-tensors $\pi$ and $\pi^{da}$,
$p\geq 3$ on $M$ are gauge equivalent, the gauge equivalence being
given by an exact $p$-form $b=da$, such that, for $t\in[0,1]$, the
function $(1 + (-1)^{p-1}tb(\pi))$ is invertible and the vector
field $\pi_t ^\sharp a$, where $\pi_t^\sharp$ is defined by $\dot{\pi}_t^\sharp=
(-1)^{p}b(\pi_t)\pi_t^\sharp,\,\,\pi_0^\sharp=\pi^\sharp$, is complete. Then there
exists a Nambu-Poisson map relating $\pi$ and $\pi^{da}$.
\end{corollary}

\begin{remark}
Although the Nambu-Poisson tensor $\pi_t$ does not depend on the
choice of the primitive $a$, the flow $\phi_t$ does. In order to
indicate the dependence of the flow $\phi_t$ on $a$ explicitly, we
will use the notation  $\phi_t^a$ for it.
\end{remark}

\begin{proposition}
For a $(p-2)$-form $\lambda$, the flow  $\phi_t^{a +
d\lambda}(\phi_t^a)^{-1}$ is generated by a Hamiltonian vector field
$X_{\lambda, a}$, i.e., there exists a $(p-2)$-form $\mu_{\lambda,
a}$ such that $X_{\lambda, a}=\pi^\sharp d\mu_{\lambda, a}$.
\end{proposition}
The explicit formal series formula for $\mu_{\lambda, a}$ can be
worked out using the BCH formula. Up to the first order in $\lambda$
we have

\begin{equation}\label{Gauge}
 \mu_{\lambda, a}= \sum_k \frac{(-\mathcal{L}_ {\pi^\sharp_t a} +
\partial_t)^k(\lambda)}{(k+1)!}\Big| _{t=0} + o(\lambda^2) \,.
\end{equation}

Applications of the SW map will be discussed in \cite{JS}.

\section{Differential graded symplectic manifolds}

Here we closely follow \cite{CattaneoSch}. Another nice discussion
of the relevant material can be found in \cite{Roytenberg}, on which
\cite{CattaneoSch} is based, and in the original paper
on the AKSZ formalism \cite{AKSZ}.
\begin{definition} A differential graded (dg-) manifold $M$ is
a graded manifold equipped with a cohomological vector field $Q$,
i.e., a graded vector field of degree +1 such that $Q^2=0$.
\end{definition}

\begin{example}The basic example of a dg-manifold is
$T[1]\Sigma$, where $\Sigma$ is an ordinary manifold.  The algebra
of smooth functions on $X=T[1]\Sigma$ is isomorphic to the algebra
$(\Omega(\Sigma), \wedge)$ of differential forms. For the
cohomological vector field on $X=T[1]\Sigma$ we take the vector
field $Q_X$ corresponding, under this isomorphism, to the de Rham
differential $d$ on $(\Omega(\Sigma), \wedge)$. If we choose some
local coordinates $x^{\mu}$ on $\Sigma$ and denote the corresponding
induced odd coordinates on the fibre of $X$ by $\theta^{\mu}$, we will have
$Q_X=\theta^{\mu}\partial_\mu$.
\end{example}

\begin{definition}
A symplectic form $\omega$ of degree $k$ on a graded manifold $M$ is
a closed, non-degenerate 2-form, which is homogeneous of degree $k$.
The corresponding graded Poisson bracket $\{ \cdot, \cdot\}$ is of degree
$-k$. It is defined similarly to the non-graded case by
$\{f,g\}:=X_f g$, where $\iota_{X_f}\omega = df$, i.e.,
$\{f,g\}=\iota_{X_f}\iota_{X_g}\omega= \omega^{-1}(df, dg)$.
\end{definition}

\begin{example}
For $V$ a smooth manifold we take $T^*[1]V$. The canonical
symplectic structure $\omega$ on  $T^*[1]V$ is of degree
$|\omega|=1$. We will denote the degree 0 local coordinates on $V$
as $X^i$ and the induced degree 1 fibre coordinates on $T^*[1]V$ by
$\chi_i$. The canonical symplectic form in these coordinates is
$\omega= d\chi_i\wedge dX^i$. The potential one-form $\vartheta$,
such that $\omega= d\vartheta$, can be taken as $\vartheta = \chi_idX^i$.
\end{example}

\begin{example}
Let $V$ be a smooth manifold. Consider $T^*[p]T[1]V$, with $p$ an
integer, $p\geq 2$. The canonical symplectic structure $\omega$ on
 $T^*[p]T[1]V$ is of degree $p$. We will denote the degree
0 local coordinates on $V$ as $X^i$ and the induced degree 1 fibre
coordinates on $T[1]V$ by $\psi^i$. Dual fibre coordinates on
$T^*[p]T[1]V\to T[1]M$, of  respective degrees  $p$ and $p-1$, will
be denoted by $F_i$ and $\chi_i$.  The canonical symplectic form in
these coordinates is $\omega= dF_i\wedge dX^i + d\psi^i \wedge
d\chi_i$. The potential one-form $\vartheta$ can be taken as
$\vartheta = F_idX^i + \psi^id\chi_i$.
\end{example}

\begin{example}
Again, let $V$ be a smooth manifold. Consider
$T^*[p]((\bigwedge^{p-1})[p-1](T[1]V))$, with $p$ an
integer, $p\geq 2$. The canonical symplectic structure $\omega$ on
 $T^*[p]((\bigwedge^{p-1})[p-1](T[1]V))$ is of degree $p$.
 We will
denote the degree 0 local coordinates on $V$ as $X^i$, the induced
degree 1 fibre coordinates on $T[1]V\to V$ by $\psi^i$, the induced
degree ${p-2}$  and degree ${p-1}$ fibre coordinates on
$(\bigwedge^{p-1}T)[p-1](T[1]V)\to T[1]V$ as
$H^{I}:=H^{i_1\ldots i_{p-1}}$, with $i_1<\ldots< i_{p-1}$ and
$\eta^I:=\eta^{i_1\ldots i_{p-1}}$, $i_1<\ldots< i_{p-1}$, respectively.
Further, the dual fibre coordinates on
$T^*[p]((\bigwedge^{p-1}T)[p-1](T[1]V))\to (\bigwedge^{p-1}T)[p-1](T[1]V)$ of the
respective degrees $p-1$, $p$, 2 and 1 will be denoted by $\chi_i$,
$F_i$, $G_I:=G_{i_1\ldots i_{p-1}}$, $i_1<\ldots< i_{p-1}$, and
$A_I:=A_{i_1\ldots i_{p-1}}$, $i_1<\ldots< i_{p-1}$. The canonical
symplectic form in these coordinates is $\omega= dF_i\wedge dX^i +
d\psi^i \wedge d\chi_i+  dG_I\wedge dH^I + d\eta^I\wedge dA_I$. The
potential one-form $\vartheta$ can be taken as $\vartheta = F_idX^i
+ \psi^id\chi_i + G_IdH^I  + \eta^IdA_I$.
\end{example}

\begin{remark}
If $V$ is a graded vector space, which has only
finitely many non-zero homogenous components (all of them
finite-dimensional) then a (formal) cohomological vector field is
the same thing as an $L_\infty$-structure on $V$ \cite{AKSZ}.
\end{remark}

\begin{remark}\label{Algebroid}
If $A$ is a vector bundle over a manifold $V$,
then a cohomological vector field on $A[1]$ is the same thing as an
Lie algebroid structure on $A$ \cite{Vaintrob}.
\end{remark}

\begin{remark} A graded symplectic form $\omega$ of a non-zero
degree $k$ is exact \cite{Roytenberg1}. The symplectic potential
$\theta$ is given by the contraction $\frac{\iota_E\omega}{k}$,
where $E$ is the graded Euler vector field, i.e., the vector field
acting on a homogeneous function $f$ of degree $|f|$ as $Ef=|f|f$.
In some homogeneous coordinates $x^i$, $E = |x^i|x^i\partial_i$.
\end{remark}

\begin{definition}
A vector field $\chi$ on a graded symplectic manifold $M$ is called
symplectic, if it preserves the symplectic structure, i.e. $\mathcal{L}_\chi
\omega =0$, where $\mathcal{L}_\chi$ is the Lie derivative with respect to
$\chi$. It is Hamiltonian, if the 1-form $\iota_\chi\omega$ is
exact, i.e., $\iota_\chi\omega=dh$ for some smooth function $h$.
\end{definition}

\begin{remark}\label{isHamiltonian}
Let $\omega$ be a symplectic form of degree $k$ and $\chi$ a
symplectic vector field of degree $l$ such that $k+l\neq 0$. Then
$\chi$ is Hamiltonian $\iota_\chi\omega =
d \left( \frac{\iota_E\iota_\chi\omega}{k+l} \right)$ \cite{Roytenberg1}.
\end{remark}

\begin{definition}
A graded manifold equipped with a graded symplectic form and a
symplectic cohomological vector field $Q$ is called a symplectic dg
manifold. It follows from the above Remark \ref{isHamiltonian}, that
if the symplectic form $\omega$ on a symplectic dg manifold ${M}$ is
of degree $k\neq -1$, then the cohomological vector field $Q$  on
$\mathcal{M}$ is Hamiltonian. Let $S$ be the corresponding
Hamiltonian function, $Q=\{S,.\}$. It can be chosen to have degree
$k+1$. Then $Q^2=0$ and $k\neq -2$ implies that $S$ is a solution to
the classical master equation
\begin{equation} \label{eqnDl}
\{S, S\}=0 \,.
\end{equation}
\end{definition}

\begin{example} A symplectic cohomological vector field $Q_M$ on
$M=T^*[1]V$ (equipped with the canonical symplectic structure) can
be determined by a Poisson 2-vector on $V$, i.e. a 2-vector field
$\pi=\frac{1}{2}\pi^{ij}\partial_i\wedge\partial_j$ with vanishing
Schouten bracket $[\pi,\pi]_S$. More explicitly,
$$
Q = \pi^{ij}\chi_j\partial_{x_i} +\frac{1}{2}\partial_k \pi^{ij}\chi_i\chi_j
\partial_{\chi_k}\,.
$$
The corresponding Hamiltonian function is
$$\gamma=\frac{1}{2}\pi^{ij}\chi_i\chi_j\,.$$
\end{example}

This last example is an illustration of the following general fact
\cite{Schwarz}.
\begin{theorem} Let $(M, \omega)$ be a degree 1 graded symplectic
structure. Then it is symplectomorphic to to $T^*[1]V$ with the
canonical symplectic form, where $V$ can be chosen to be an ordinary
manifold. Under the symplectomorphism, the degree -1 Poisson bracket
$\{\cdot, \cdot\}$ on $C^\infty$ is mapped to the Schouten bracket $[\cdot, \cdot]_S$
on $T^*[1]V=\Gamma(\bigwedge TV)$. A smooth vector function $S$ of
degree $2$ is mapped to a bivector field $\pi$ on $V$. Hence, $S$
solves the classical master equation is translated into  $\pi$
being Poisson.
\end{theorem}

\begin{remark} The above Theorem can be restated as follows.
A graded symplectic manifold of non-negative
degree is called a $N$-manifold if all its coordinates are of
non-negative degree. Isomorphism classes of dg symplectic
$N$-manifolds of degree 1 are one-to-one with isomorphism classes of
Poisson manifolds.
\end{remark}

\begin{remark} It is also known (letter 7 of  \cite{Severa7} and \cite{Roytenberg1})
that isomorphism classes of dg symplectic
$N$-manifolds of degree 2 are one-to-one with isomorphism classes of
Courant algebroids.
\end{remark}

\begin{example}\label{TwDorf}
In the example of $T^*[p]T[1]V$, we have
the canonical symplectic cohomological vector field
$\psi^i\partial_{X^i}$, corresponding to the de Rham differential
on $V$.  The corresponding degree $(p+1)$ Hamiltonian function is
$-\psi^iF_i$. Hence, if $c$ is a closed $(p+1)$-form, the sum
$$S=-\psi^iF_i +\frac{1}{(p+1)!}{c}_{i_1\ldots
i_{p+1}}{\psi}^{i_1}\ldots {\psi}^{i{p+1}}$$ is also a solution to
the master equation and the corresponding Hamiltonian vector field
$Q=\{S,\cdot\}$ is another example of a symplectic cohomological vector
field on $T^*[p]T[1]V$.
\end{example}

\subsubsection*{Twisted Dorfman bracket}

Using the above cohomological vector field $Q$ on $M=T^*[p]T[1]V$, the
twisted Dorfman bracket can be identified as a derived bracket on
$T^*[p]T[1]V$. Let $C_n$ for $n \leq p-1$ denote the  subspace of
all degree $n$ functions on $T^*[p]T[1]V$. For $n \leq p-2$, a
degree $n$ function $\tilde\alpha\in C_n$ corresponds to an $n$-form
$\alpha \in \Omega^n(V)$ via
$\tilde\alpha=\frac{1}{(n)!}\alpha_{i_1\ldots
i_{n}}{\psi}^{i_1}\ldots {\psi}^{i_{n}}$. Also, a degree $p-1$
function $\tilde e= \frac{1}{(p-1)!}\omega_{i_1\ldots
i_{p-1}}{\psi}^{i_1}\ldots {\psi}^{i{p-1}}+ v^i\chi_i\in C_{p-1}$ on
$T^*[p]T[1]V$ corresponds to a pair $e=(\omega, v)\in
\Omega^{p-1}(V) \oplus \mathfrak X(V)$ consisting of a $(p-1)$-form and a
vector field. We have the following relation between the dg
symplectic manifold structure of $T^*[p]T[1]V$ and the Courant
algebroid structure on $\mathfrak X(V) \oplus \Omega^{p-1}(V)$ (given by
the pairing, the anchor, the twisted Dorfman bracket and the
differential as described in Section \ref{SectionDorfman}.)
$$\{\tilde e_1, \tilde e_2\}={\langle e_1, e_2\rangle}^\sim \,,$$
$$\{\{S,\tilde e\},f\}=\rho(e)f\,,$$
$$\{\{S,\tilde e_1\}, \tilde e_2\}=[e_1,e_2]^\sim_c\,,$$
$$\{S,\tilde \alpha\}=(d\alpha)^\sim\,,$$
where, $\alpha$ is $(p-2)$-form and $f$ is a function on $M$.

In particular, the subspace of degree $p-1$ functions on
$T^*[p]T[1]V$ is closed under the derived bracket $\{\{S,.\},.\}$,
which can be identified with the twisted Dorfman bracket. Actually,
as a consequence of a theorem in \cite{Getzler}, the complex
$(C_n[p-1],\psi_i\partial_{\chi_i})_{n=0}^{p-1}$ can be equipped
with an Lie $p$-algebra structure\footnote{See \cite{Zambon} for the
explicit formulas. For the original work on case $p=2$, see
\cite{RoytenbergWeinstein}. Also, see \cite{Bering} for another
related (generalized) $L_\infty$-structure.}, the twisted Dorfman
bracket being one of the binary brackets.

Let us note that in the case of $p=2$ the canonical transformation
$e^{\delta_\zeta}$ generated by a degree $2$ function
$\zeta=\zeta^{ij}\chi_i\chi_j$ gives $e^{\delta_\zeta} \tilde e=
(e^\zeta e)^\sim$.

\subsubsection*{Twisted Dorfman bracket -- continuation}\label{DorfmanRemark}

In the example of  $M=T^*[p]((\wedge^\bullet T)[p-1](T[1]V))$,
we can take as the Hamiltonian vector field
$Q=\{S,\cdot \}$ corresponding to the Hamiltonian\footnote{We use the
following convention. Only when using the upper case notation for a
multi-index, we assume it ordered, otherwise not.}
$$S = -\psi^iF_i + G_I\eta^I +\frac{1}{(p+1)!}{c}_{i_1\ldots
i_{p+1}}{\psi}^{i_1}\ldots {\psi}^{i{p+1}}\,.$$ We can embed
$T^*[p]T[1]V$ into $T^*[p]((\wedge^\bullet T)[p-1](T[1]V))$ as the zero
section of the vector bundle $T^*[p]((\wedge^\bullet T)[p-1](T[1]V))\to
T^*[p]T[1]V$. Obviously, this embedding is a Poisson map. Hence, the
Dorfman bracket can be identified as (a part of) the restriction to
$T^*[p]T[1]V$ of the derived bracket on
$T^*[p]((\wedge^\bullet T)[p-1](T[1]V))$ too. For the further reference,
let us also note that under the canonical transformation generated
by the degree $p$-function $-\frac{1}{(p-1)!}\psi^{i_1}\ldots
\psi^{i_{p-1}}A_{i_1\ldots i_{p-1}}$ the Hamiltonian function $S$
changes to
$$S' = -\psi^iF_i + \frac{1}{(p-1)!}G_{{i_1}\ldots {i_{p-1}}}(\eta^{{i_1}
\ldots {i_{p-1}}} - \psi_{i_1}\ldots \psi_{i_{p-1}})+\frac{1}{(p+1)!}{c}_{i_1\ldots
i_{p+1}}{\psi}^{i_1}\ldots {\psi}^{i{p+1}}\,.$$ This gives another way
of identifying the Dorfman bracket as (a part of) the restriction of
the derived bracket on $T^*[p]((\wedge^\bullet T)[p-1](T[1]V))$.

\section{AKSZ construction}

In this section we review the AKSZ construction.
We follow mainly \cite{CattaneoSch} and \cite{HofmanPark},
where the interested reader can find missing details. See also
\cite{Roytenberg}, \cite{CattaneoFelderAKSZ} \cite{Park} and, of
course, \cite{AKSZ}. The AKSZ formalism \cite{AKSZ} provides for a
solution of the classical master equation \eqref{eqnDl} on the space of maps
$\mathcal{M}:=C^{\infty}(X,M)=M^X$.

Here,
\begin{itemize}
\item[(i)] the source $(X, Q_X, \mu)$ (a super worldvolume) is a
differential graded manifold $X$, equipped with a measure $\mu$ which
is invariant under the cohomological vector field $Q_X$, and
\item[(ii)]  the target $(M,Q_M,\omega)$ (a super spacetime) is a
differential graded symplectic manifold $M$  with the graded
symplectic form $\omega$, such that the cohomological vector field
$Q_M$ is a Hamiltonian vector field.
\end{itemize}
Using the above structures on $X$ and $M$ the AKSZ construction produces:
\begin{itemize}
\item[(i)] a graded symplectic structure $\check{{\omega}}$ on
the space of maps $\mathcal{M}$, and
\item[(ii)] a symplectic cohomological vector field
${Q}$ on $\mathcal{M}$.\\
\end{itemize}

\subsubsection*{The cohomological vector field on the space of maps}

We describe very briefly the construction. The tangent space $T_f\mathcal{M}$ to
$\mathcal{M}$, at some function $f\in \mathcal{M}$,
is identified with with the space of sections $\Gamma(X, f^*TM)$.
Then a vector field on $\mathcal{M}$ is an assignment of an element
in $T_{f(x)}M$ to each $x\in X$ and $f\in \mathcal{M}$. In
particular, the vector fields $Q_0$ and $\check Q$ on $\mathcal{M}$,
associated to the cohomological vector fields $Q_X$ and
$Q_M$, respectively, are defined as
$$
(Q_0f)(x)= Q_0(x,f)= df(x)Q_X(x) \,,
$$
and
$$(\check Q f)(x)=\check Q(x,f) = Q_M(f(x))\,.
$$
We note that $Q_0$ and $\check Q$  are of the degree $1$,
square to zero and graded commute with each other.
\begin{proposition}   Let $X$ and $M$ be differential graded manifolds. Then the space
of smooth maps $\mathcal{M}=M^X$ is a differential graded manifold,
with cohomological vector field
$$Q=Q_0 + \check Q\,.$$
\end{proposition}

\subsubsection*{The source}

For our purposes it will be sufficient to
consider the case $X=T[1]\Sigma$, where $\Sigma$ is an ordinary
manifold of dimension $p+1$ with boundary $\partial\Sigma$. The
algebra of smooth functions on $X=T[1]\Sigma$ is isomorphic to the
algebra $(\Omega(\Sigma), \wedge)$ of differential forms. We
denote the isomorphism $j$. For the cohomological vector field on
$X=T[1]\Sigma$ we take the vector field $Q_X$ corresponding, under
this isomorphism, to the de Rham differential $d$ on
$(\Omega(\Sigma), \wedge)$. If we choose some local coordinates
$x^{\mu}$ on $\Sigma$ and denote the corresponding induced odd
coordinates on the fibre by $\theta^{\mu}$,\footnote{Hence, a point
$x\in X$ is locally parametrized by $(x^{\mu}, \theta^{\mu})$.} we
will have $Q_X=\theta^{\mu}\partial_\mu$. Take the canonical measure
$\mu$ on $X=T[1]\Sigma$, which maps a function $f$ on $X$ to $\int_X
f :=\int_\Sigma j(f)$, where $j$ is the isomorphism between smooth
functions on $T[1]\Sigma$ and smooth forms $\Omega(\Sigma)$ on
$\Sigma$.  In local coordinates $\mu= d^{p+1}x\, d^{p+1}\theta$.

\subsubsection*{The graded symplectic structure on $\mathcal{M}$}

For any $n$-form $\alpha\in \Omega^n(M)$, we obtain an $n$-form
$\check {\alpha}\in \Omega^n(\mathcal{M})$ by
\begin{equation} \label{eqnEa}
\check{\alpha}= \int_X \mbox{ev}^* (\alpha) \,,
\end{equation}
where $\mbox{ev}^* $ is the pullback under the evaluation map
$\mbox{ev}:X\times\mathcal{M} \to M$, defined by
$\textrm{ev}(x,f) = f(x)$. Integrating over
$X=T[1]\Sigma$ we obtain an $n$-form of degree $|\alpha|-(p+1)$. In
particular, from the symplectic form $\omega$ on $M$, we get the
symplectic form $\check \omega$ on $\mathcal{M}$ and if there exists
a symplectic potential $\vartheta$ on $M$, we obtain a
corresponding symplectic potential $\check \vartheta$ on
$\mathcal{M}$ as well.
For example, from a degree $p+1$ function $f\in \Omega^0(M) = C^\infty(M)$,
we obtain a function $\check f\in\Omega^0(\mathcal M) = C^\infty(\mathcal M)$,
defined on $\phi\in \mathcal M$ by
$$
\check f [\phi] =  \int_X \phi^*(f) \,.
$$
Furthermore, we can use the coordinates on $M$, say $X^i$, to parametrize a general the
superfield $\phi: X\to M$, hence introduce the ``coordinate" superfields\footnote{In general,
we may expand a superfield $\phi: X\to M$
as a polynomial in the odd variables $\theta$ with coefficients
being $M$-valued functions on $\Sigma$, i.e., $\phi(x^\mu,
\theta^\mu)= \phi^{(0)}(x^\mu) + \phi^{(1)}_{\nu}(x^\mu)\theta^\nu +
\ldots + \phi^{(p+1)}_{\nu_1\ldots
\nu_{p+1}}(x^\mu)\theta^{\nu_1\ldots \nu_{p+1}}$.  Then if
$\phi^{(0)}$ is degree $k$ $\phi^{(i)}$ will be degree $k-i$. Assume
that $M$ is non-negatively graded. Factoring out the ideal generated
by coefficients with negative degrees will give the space of the
fields (including the ghosts) and factoring out the ideal generated
by coefficients with nonzero degrees (ghost and antifields) will
give the space of classical fields.}
$$
\phi^i (x^\mu)=\phi^*(X^i)(x^\mu) \,.
$$
In particular, in the case
$|\omega|=p$ corresponding to the BV formalism, which we will
consider from now, we have
\begin{proposition}
For a degree $p$ symplectic form $\omega$ on $M$, the 2-form
$\check{\omega}$ is a degree -1 symplectic form on $\mathcal{M}$.
Further, if $\vartheta$ is a symplectic potential for $\omega$, then
$\check{\vartheta}$ is a symplectic potential for $\check{\omega}$.
Moreover, since $\iota_{\check Q}\, \check{\omega}= \int_X \operatorname{ev}^*
\iota_{Q_M}\omega$, we  also have $\mathcal{L}_{\check Q}\check{\omega}=0$. In
particular, if $\gamma$ is the a degree $p+1$ Hamiltonian function
on $M$ corresponding to $Q$ then $\check S := \check{\gamma}=\int_X
\operatorname{ev}^* \gamma$ is the degree 0 Hamiltonian function on
$\mathcal{M}$ corresponding to $\check Q$
\end{proposition}
Let $\boldsymbol\{\cdot,\cdot\boldsymbol\}$ be the degree 1 Poisson bracket,
the BV bracket, on $\mathcal{M}$ corresponding to the degree $-p$
Poisson bracket $\{\cdot,\cdot\}$ on $M$.
\begin{proposition}\label{PoissonMap}
The map $\int_X \operatorname{ev}^*$ is a degree $-(p+1)$ Lie algebra map
from $(M,\{\cdot,\cdot\})$ to $(\mathcal{M},\boldsymbol\{\cdot,\cdot\boldsymbol\})$,
i.e., $\int_X \operatorname{ev}^*\{f,g\}=\boldsymbol\{\int_X
\operatorname{ev}^*f,\int_X \operatorname{ev}^*g\boldsymbol\}$, for any two
functions $f,g$ on $M$.
\end{proposition}

\noindent{\bf{Notation}} We will use the following notation. For a
function $f$ on $M$, we  will omit the $\textrm{ev}^*$ symbol under
the integral sign and simply write  $\int_X f$ instead of $\int_X \mbox{ev}^*f$, etc.

\subsubsection*{Solution to the master equation}

Since $p\geq 0$, we will assume that we have chosen a symplectic
potential $\vartheta$ on $M$ and that $Q_M$ is Hamiltonian with the
degree $(p+1)$ Hamiltonian function $\gamma$. To proceed further, we
should be careful about the boundary conditions. This is discussed
in great detail in \cite{HofmanPark}, \cite{HofmanPark1} and
\cite{CattaneoFelderAKSZ}. We will need a slight modification of
that discussion. Let $L$ denote the Lagrangian submanifold of $M$
which is the zero locus of $\vartheta$, and $L'\subset L$ some
submanifold. We will consider only a subspace $\mathcal{M}_{L'}$ of
$\mathcal{M}$, which consists of maps that map the boundary
$\partial X = T[1]\partial \Sigma$ into $L'$.\footnote{To be more
precise, $\mathcal{M}_{L'}$ has to be properly regularized, see
\cite{CattaneoFelderAKSZ} for the detailed discussion for $L'=L$. We
will not discuss the subtleties related to this and refer to
\cite{CattaneoFelderAKSZ} for the general discussion, cf. also
\cite{HofmanPark1}.} Also, we will assume that $\gamma$ when
restricted to $L'$ vanishes. Hence, we assume
$$\{\gamma,\gamma\}=0\,,$$ and $$\gamma|_{L'}=0\,.$$

\begin{theorem}\label{solution Master}On
$\mathcal{M}_L$, the vector field $Q_0$ is Hamiltonian, with the
corresponding Hamiltonian function
$$S_0= -\iota_{Q_0}\check{\vartheta}\,.$$
For $\gamma$ satisfying $\{\gamma,\gamma\}=0$,
and $\gamma|_{L'}=0$ the sum
$$S= S_0 + \check \gamma\,,$$ is a solution of the master equation on
$\mathcal{M}_{L'}$, i.e., $S$ is a BV action.\footnote{I.e., for $\phi\in
\mathcal{M}=M^X$, $S[\phi]=  S_0[\phi] + \check \gamma[\phi]=
\int_X (-\iota_{Q_X}\phi^*\alpha +  \phi^*(\gamma))$.}
\end{theorem}

We will not give a formal proof, since it is a only a slight
modification of the discussion, in the case $L'=L$ in
\cite{CattaneoFelderAKSZ} and \cite{HofmanPark}. We sketch as an
example the case when $X=T[1]\Sigma$ and $M=T^*[p]N$, with $N$ some graded
manifold $N$, with coordinates of degrees from 0 to $p$. Choose
homogeneous local coordinates $X^i$ on $N$ and the dual fibre
coordinates $P_i$ on $T^*[p]N\to N$. For the symplectic potential on
$M$ we take
$$\vartheta = P_i\, dX^i\,.$$
Hence, $L$ is given by $P=0$.

We will use the notation $\boldsymbol{X^i},\boldsymbol{P_i}$ for the
superfields associated with the local coordinates $X^i,P_i$ on $M$
respectively, i.e.,
$\boldsymbol{X}^i=\Phi^*X^i,\,\boldsymbol{P}_i=\Phi^*P^i$.

So in the local coordinates
$$\check\omega =\int_X \delta \boldsymbol{P}_i\wedge \delta \boldsymbol{X}^i\,,$$ and
the BV bracket is determined by the bivector\footnote{Here the superscripts
$R$ and $L$ refer the right and left
(functional) derivative. We will omit them in the sequel.}
$$
\int_X \partial^R_{\boldsymbol{{X}}^i}\wedge
\partial^L_{\boldsymbol{{P}}_i} \,.
$$
Hence, for the $S_0$ part of the BV action we have\footnote{More precisely, we should have
written $S_0[\phi]$ instead of $S_0$. We will continue, hopefully
without causing confusion, with this shorthand notation in the
sequel.\label{notationphi}}
$$
S_0=\int_X {\boldsymbol{P}}_i\, {D\boldsymbol{X}}^i \,.
$$ In the above formulas and in the
sequel we use the notation $D$ for $\theta^\mu\partial_\mu$. Now we
can explicitly check that $S_0$ is the Hamiltonian for $Q_0$ iff
$\int_XD(\boldsymbol{P}_iD\boldsymbol{X}^i)=\int_{\partial
X}\boldsymbol{P}_iD\boldsymbol{X}^i=0$. By assumption,
$(\boldsymbol{P}_i(x), \boldsymbol{X}^i(x))\in L'\subset L$ on the
boundary and it follows that for $x\in \partial X$ we have
$\boldsymbol{P}^i(x)=0$. Therefore, $S_0$ is indeed the Hamiltonian
for $Q_0$.

Furthermore, we have
$$Q_0 \check \gamma = \int_X D{\gamma} = \int_{\partial X}{\gamma}\,. $$
where we used the symbol $D$ also for the lift of
$\theta^\mu\partial_\mu$ to $\mathcal{M}\times X$. For $x\in
\partial X$, by assumption, $(\boldsymbol{P}_i(x),
\boldsymbol{X}^i(x))\in L'$ and we see that if $\gamma|_{L'}$
vanishes then $Q_0 \check \gamma=\{S_0, \check\gamma\}=0$. We
conclude that $S_0+\check \gamma$ is a BV action.

\begin{remark}\label{dgmaps}
Integrating in $S$ over the odd variables
$\theta$ and restricting it to the degree zero fields, we obtain the
``classical" action $S_{cl}$. Then the solutions of the classical
field equations are dg maps from $(T[1]\Sigma, D)$ to $(M,
(-1)^{p+1}Q)$  \cite{Roytenberg}. This follows from the fact that the critical points of
$S$ are the fixed points of $Q$.
\end{remark}

\subsubsection*{Canonical transformations}

{}From the above discussion,
it follows that a canonical transformation on $M$, generated by a
function $\alpha$ of degree $p$, induces a canonical transformation
on $\mathcal{M}$ generated by the  function $\check\alpha$. We will
use the notation $\delta_{\alpha}$ for the corresponding Hamiltonian
vector field and $e^{\delta_{\alpha}}$ and $e^{\delta_{\check
\alpha}}$ for the respective canonical transformations. From
$e^{-\delta_\alpha}\{e^{\delta_\alpha}\gamma,e^{\delta_\alpha}\gamma\}=
\{\gamma, \gamma\}$ we see that
$\{e^{\delta_\alpha}\gamma,e^{\delta_\alpha}\gamma\}=0$, provided
$\{\gamma,\gamma\}=0$. We can write
$$e^{\delta_{\check \alpha}}(S_0 + \int_X
\gamma )=S_0 + \int_X e^{\delta_\alpha}\gamma + \sum_{n\geq
1}\frac{1}{n!}\delta_{\check \alpha}^{n-1}\int_{\partial X}\alpha \,.$$
Hence, when $\alpha|_{L'}=0$, the BV actions $S_0 + \int_X \gamma $
and $S_0 + \int_X e^{\delta_\alpha}\gamma$ are equivalent, i.e.,
related by a canonical transformation.

In general, the symplectic potential $\vartheta$ (and hence also the
the Lagrangian submanifold $L$ defined by its locus) or the
submanifold $L'$ may have changed due to the canonical
transformation. Hence, for degree $p$ generating function $\beta$ on
$M$ it may happen that $L'_\beta:=e^{\delta_\beta}(L') \neq L'$. So,
let us assume that $(e^{\delta_\beta}\gamma)|_{L'}=0$ and,
therefore, $S_0 + (e^{\delta_\beta}\gamma){\check\,}$ is a BV action.
Also assume, for simplicity, that $\{\beta,\beta\}=0$. We have
\begin{align} S_0 + \int_X e^{\delta_\beta}\gamma &\sim
e^{-\delta_{\check \beta}}(S_0 + \int_X e^{\delta_\beta}\gamma )=
e^{-\delta_{\check \beta}}S_0 + \int_X
e^{-\delta_{\beta}}e^{\delta_\beta}\gamma \nonumber\\&= S_0 - \int_X D\beta
+ \int_X \gamma = S_0 + \int_X \gamma - \int_{\partial X}\beta\,, \nonumber
\end{align}
where in the first equality we have used Proposition \ref{PoissonMap} and in
the second equality the fact that only first two terms in the
expansion of $ e^{-\delta_{\check \beta}}$ will survive due to the
(graded) Jacobi identity and the assumption  $\{\beta,\beta\}=0$.

Hence, we have a slight modification of the corresponding statement
of \cite{HofmanPark}

\begin{theorem}\label{actionEquiv} \mbox{}
\begin{itemize}
\item[(i)] Assume that $\{\gamma,\gamma\}=0$, $\gamma|_{L'}=0$ and
$\alpha|_{L'}=0$. The BV actions $S_0 + \int_X \gamma $ and $S_0 +
\int_X e^{\delta_\alpha}\gamma$ are equivalent, i.e., related by a
canonical transformation.
\item[(ii)] Assume that $\{\gamma,\gamma\}=0$ and $\{\beta,\beta\}=0$. Also,
assume that $(e^{\delta_\beta}\gamma)|_{L'}=0$. Then the BV action
$S_0 + \int_X e^{\delta_\beta}\gamma$ on $\mathcal{M}_{L'}$ and the
BV action $S_0^\beta + \int_X\gamma$ on $\mathcal{M}_{L'_\beta}$,
where $S_0^\beta$ corresponds to the symplectic potential $\vartheta
- d\beta$, are equivalent. The latter BV action can be written as a
bulk/boundary action $S_0  + \int_X\gamma - \int_{\partial X}\beta$.
\end{itemize}
\end{theorem}

Related to this we have the following corollary \cite{HofmanPark}
\begin{corollary}\label{MainGaugeTr}Let $\alpha$ and $\beta$ be
degree $p$ functions on $M$, $\alpha|_{L'}= 0$ and $\beta \in {\rm Im}\, i_{L'}$.
Here $i_{L'}$ is some embedding of functions on $L'$ into functions on
$M$ such that $P_{L'} i_{L'} = \rm{id}$  with
$P_{L'}$ being the restriction to $L'$. Assume that we have found
$\alpha'|_{L'} = 0$,  and $\beta'\in {\rm Im}i_{L'}$ such that
$e^{\delta_\alpha} e^{\delta_\beta}=
e^{\delta_{\beta'}}e^{\delta_{\alpha'}}$. Then the BV actions $S_0 +
\int_X \gamma + \int_{\partial X}\beta$ and $S_0 + \int_X
e^{\delta_\alpha }\gamma +\int_{\partial X}\beta'$ are
equivalent.\footnote{Of course, we assume that $\{\gamma,\gamma\}=0$,
$\gamma|_{L'_\beta}=0, \{\beta,\beta\}=0$ and $\{\beta',\beta'\}=0$.}
\end{corollary}
The statement of the corollary follows from the chain of equivalences and equalities
\begin{align*} S_0 + \int_X \gamma + \int_{\partial X}\beta &\sim  S_0
+ \int_X e^{-\delta_\beta}\gamma
\sim S_0  + \int_X e^{\delta{_\alpha'}} e^{-\delta_\beta}\gamma \\ &=
S_0  + \int_X e^{-\delta_{\beta'}} e^{\delta_\alpha} \gamma \sim
S_0  + \int_X e^{\delta_\alpha} \gamma + \int_{\partial X}\beta' \,,
\end{align*}
where we have used that, according to the assumptions,
$$  e^{\delta_\alpha}e^{-\delta_{\beta'}}\gamma|_{L'}=
e^{\delta_{\alpha'}} e^{-\delta_\beta}\gamma|_{L'}=e^{-\delta_\beta}\gamma|_{L'}=0.$$

\section{Examples of the AKSZ construction}

\subsection{Poisson sigma model}

Here we follow \cite{CattaneoFelderAKSZ}. For $V$ a smooth manifold,
we take $M=T^*[1]V$. The canonical symplectic structure $\omega$ on
the target $T^*[1]V$ is of degree $|\omega|=1$. Hence, we take a
$2$-dimensional $\Sigma$. We will denote the degree 0 local
coordinates on $V$ as $X^i$ and the induced degree 1 fibre
coordinates on $T^*[1]V$ by $\chi_i$. The canonical symplectic form
in these coordinates is $\omega= d\chi_i\wedge dX^i$. The potential
one-form $\vartheta$ can be taken as $\vartheta = \chi_i\, dX^i$. Its
zero locus $L$ is given by $\chi_i=0$.  A symplectic cohomological
vector field $Q_M$ on $M=T^*[1]V$ is necessarily determined by a
Poisson 2-vector
 on $V$, i.e. a
2-vector  field $\pi=\frac{1}{2}\pi^{ij}\partial_{X^i}\wedge\partial_{X^j}$
with vanishing Schouten bracket $[\pi,\pi]_S$. More explicitly,
$$Q_M = \pi^{ij}\chi_j\partial_{X_i} +\frac{1}{2}(\partial_k \pi^{ij})\chi_i\chi_j
\partial_{\chi_k}\,.$$
The corresponding Hamiltonian function is
$$\gamma=\frac{1}{2}\pi^{ij}\chi_i\chi_j\,.$$
Furthermore, we may use the superfields
$\boldsymbol{X}^i,\boldsymbol{\chi}^i$ to write the BV bracket in
the form
$$\int_X \partial_{{\boldsymbol{X}}^i}\wedge
\partial_{{\boldsymbol{\chi}}_i} \,.$$
For the $S_0$ part of the BV action we have
$$S_0=\int_X {\boldsymbol{\chi}}_i{D\boldsymbol{X}}^i\,.$$
Hence, in this case, the AKSZ construction gives the Poisson sigma
model of \cite{Ikeda}, \cite{StroblSchaller}.\footnote{Recall
footnote \ref{notationphi}.}
$$S=\int_X ({\boldsymbol{\chi}}_i{D\boldsymbol{X}}^i +\frac{1}{2}
\boldsymbol{\pi}^{ij}\boldsymbol{\chi}_i\boldsymbol{\chi}_j)\,.$$ It follows from Remarks
\ref{dgmaps} and \ref{Algebroid} that the solutions to the classical
field equations are Lie algebroid maps from $T[1]\Sigma$ to
$T^*[1]V$, cf.\ \cite{SeveraSome}.

If the boundary $\partial\Sigma$ is nonzero, we take the Dirichlet
boundary conditions for the superfields $\boldsymbol{\chi}_i$.

\subsection{Open topological membrane and twisted Poisson sigma
model}

This subsection is based on \cite{HofmanPark}. Let $V$ be a
smooth manifold and put $M=T^*[p]T[1]V$, with $p$ an integer $p\geq
2$. The case $p=2$, corresponds to the open topological membrane.
The canonical symplectic structure $\omega$ on the target
$T^*[p]T[1]V$ is of degree $p$. Hence, we take a $p+1$ dimensional
$\Sigma$. We will denote the degree 0 local coordinates on $V$ as
$X^i$ and the induced degree 1 fibre coordinates on $T[1]V$ by
$\psi^i$. Dual fibre coordinates on $T^*[p]T[1]V\to T[1]V$ of the
respective degrees $p-1$ and  $p$ will be denoted by $\chi_i$ and
$F_i$.  The canonical symplectic form in these coordinates is
$\omega= dF_i\wedge dX^i + d\psi^i \wedge d\chi_i$. The potential
one-form $\vartheta$ can be taken as $\vartheta = F_idX^i +
\psi^id\chi_i$. Its zero locus $L$ is given by $F_i=0$ and
$\psi^i=0$.

We may use the superfields corresponding to the local coordinates on $T^*[p]T[1]V$ to write the BV
bracket in the form
$$\int_X (\partial_{{\boldsymbol{X}}^i}\wedge
\partial_{{\boldsymbol{F}}_i}
+\partial_{{\boldsymbol{\chi}}_i}\wedge
\partial_{{\boldsymbol{\psi}}^i}) \,.$$
For the $S_0$ part of the BV action we have
$$S_0=\int_X ({\boldsymbol{F}}_i{D\boldsymbol{X}}^i +
{\boldsymbol{\psi}}^i{D\boldsymbol{\chi}}_i)\,.$$

Furthermore, the canonical symplectic cohomological vector field
corresponding to the de Rham differential on $V$ is
$Q=\psi_i\partial_{X_i}$ and the corresponding degree $(p+1)$
Hamiltonian $\gamma_0=\psi^iF_i$ is a solution to the classical
master equation on $T^*[p]T[1]V$ satisfying $\gamma_0|_L=0$. From
now on, we will use notation $\Gamma_0$ for $\check \gamma_0$.
Hence we have the ``free part'' of the BV action
$$S_0+ \Gamma_0 =\int_X ({\boldsymbol{F}}_i{D\boldsymbol{X}}^i +
{\boldsymbol{\psi}}^i{D\boldsymbol{\chi}}_i - \boldsymbol{\psi}^i
\boldsymbol{F}_i)\,.$$

\subsubsection*{Bulk interaction}

Let $c$ be a $(p+1)$-form on $V$. In
local coordinates,
$$c=\frac{1}{(p+1)!}c_{i_1\ldots i_{p+1}}\, dX^{i_1}\wedge \ldots \wedge dX^{i_{p+1}}\,.$$
We associate with it the degree $p+1$ function
$C=\frac{1}{(p+1)!}{c}_{i_1\ldots i_{p+1}}{\psi}^{i_1} \ldots
{\psi}^{i{p+1}}$ on $M$ and the corresponding degree $0$ function
$\Gamma_1=\int_X C$ on $\mathcal M$. By construction,
$$\boldsymbol{\{}S_0 + \Gamma_0 + \Gamma_1, S_0 +
\Gamma_0 + \Gamma_1, \boldsymbol{\}} =2  \boldsymbol{\{}\Gamma_0,
\Gamma_1, \boldsymbol{\}}=2 \int_X \{-\boldsymbol{\psi}^i
\boldsymbol{F}_i, \boldsymbol{C}\}\,.$$
Since the function $-\psi^i
F_i$ on $T^*[p]V$ corresponds to the de Rham differential on $V$, we
see that the sum
$$S_0 + \Gamma_0 + \Gamma_1 = \int_X ({\boldsymbol{F}}_i{D\boldsymbol{X}}^i +
{\boldsymbol{\psi}}^i{D\boldsymbol{\chi}}_i - \boldsymbol{\psi}^i
\boldsymbol{F}_i + \frac{1}{(p+1)!}\boldsymbol{c}_{i_1\ldots
i_{p+1}}{\boldsymbol{\psi}}^{i_1}\ldots
{\boldsymbol{\psi}}^{i{p+1}})$$ is a solution to the master equation
iff the $(p+1)$-form $c$ is closed. Let us also note, that the
canonical transformation on $M$ generated by the degree $p$ function
${B}$, where $B= \frac{1}{p!}{b}_{i_1\ldots i_{p}}{\psi}^{i_1}
\ldots {\psi}^{i_p}$ corresponds to a $p$-form $b$ on $V$, amounts to
the gauge transformation $c\mapsto c - db$.

\subsubsection*{Boundary interaction}

One can also  consider $\Sigma$
to have a nonempty boundary $\partial \Sigma$. For our discussion it
is relevant that the boundary conditions can be chosen so that the
superfields $\boldsymbol{X}^i$, $\boldsymbol{F}_i$,
$\boldsymbol{\psi}^i$ and $\boldsymbol{\chi}_i$ restrict on the
boundary to maps to the zero locus of $\vartheta$. This means that
we take the Dirichlet boundary conditions for the superfields
$\boldsymbol{F}_i$ and $\boldsymbol{\psi}^i$.\footnote{Other
boundary conditions and the related effects are discussed in
\cite{HofmanPark}. Also, see \cite{HofmanPark1} and
\cite{CattaneoFelderAKSZ} for discussion of boundary conditions for
the remaining superfields.}

The case $p=2$ was thoroughly discussed in \cite{HofmanPark}. If
$p=2$,  we can consider a boundary term associated to a 2-vector
field $\pi$ on $V$. This can be done considering the canonical
transformation on $M$ generated by the degree 2 function
$-\frac{1}{2}{\pi}^{ij}{\chi}_i{\chi}_j$. We know that the result is
equivalent to the boundary/bulk BV action

$$S = \int_X ({\boldsymbol{F}}_i{D\boldsymbol{X}}^i +
{\boldsymbol{\psi}}^i{D\boldsymbol{\chi}}_i - \boldsymbol{\psi}^i
\boldsymbol{F}_i +
\frac{1}{6}\boldsymbol{c}_{ijk}\boldsymbol{\psi}^{i}\boldsymbol{\psi}^{j}\boldsymbol{\psi}^{k})
+ \frac{1}{2}\int_{\partial X}
\boldsymbol{\pi}^{ij}\boldsymbol{\chi}_i\boldsymbol{\chi}_j\,.$$ For
$\gamma=\psi^i F_i +
\frac{1}{6}{c}_{ijk}{\psi}^{i}{\psi}^{j}{\psi}^{k}$, the condition
$\{\gamma, \gamma\}=0$ gives, as before, $dc=0$. Regarding the
condition $\gamma|_{L_\pi}=0$, we notice that the Lagrangian
submanifold ${L_\pi}$ is given by equations
$$\psi^i= \{\frac{1}{2}{\pi}^{ij}{\chi}_i{\chi}_j, \psi^i\}=
\pi^{ij}\chi_j\,,$$ and
$$F_i=\{\frac{1}{2}{\pi}^{ij}{\chi}_i{\chi}_j, F_i\}=
\frac{1}{2}\partial_i\pi^{jk}\chi_j\chi_k\,.$$
Let us note that the first equation gives the graph of the map $\pi$,
whereas the second equation is the integrability condition for this graph.
These two conditions determine a Dirac structure in $T^*V\oplus TV$.\footnote{The paper
\cite{HofmanPark} discusses also the more general case of $M=T^*[2]T[1]A$, where $A$ is a Lie
algebroid, and describes the correspondence between Lagrangian submanifolds on $M$ and
Dirac structures in $A^*\oplus A$ in this more general case. }

{}From here if follows that
$$[\pi,\pi]_S=-\wedge^3\pi^\sharp c\,,$$
where the bracket subscript $S$ stands for the Schouten bracket.
Hence, $\pi$ defines a twisted Poisson bracket on V. This is, up to
equivalence, the most general BV action of the AKSZ form in our case
\cite{HofmanPark}.

\subsubsection*{Gauge transformations}

Recall Corollary \ref{MainGaugeTr}. Following \cite{HofmanPark}, we
consider a degree 2 function
$\alpha=\frac{1}{2}\alpha_{ij}\psi^i\psi^j$, corresponding to a
2-form $\tilde\alpha=\frac{1}{2}\alpha_{ij}dX^i\wedge dX^j$ on $V$
and a degree 2 function $\beta=\frac{1}{2}\pi^{ij}\chi_i\chi_j$
corresponding to a twisted Poisson tensor
$\pi=\frac{1}{2}\pi^{ij}\partial_i\wedge\partial_j$,
$[\pi,\pi]_S=-\wedge^3\pi^\sharp c$ on $V$. From Proposition
\ref{1.2} it follows that in Corollary \ref{MainGaugeTr} we have to
choose $\beta'=\frac{1}{2}\pi'^{ij}\chi_i\chi_j$, with
${\pi'}^\sharp=\pi^\sharp\circ(1+\alpha_\sharp\circ\pi^\sharp)^{-1}$,
$[\pi',\pi']_S=-\wedge^3\pi'^\sharp(c-d\alpha)$.\footnote{We will
describe the original argument of \cite{HofmanPark} later in
relation with the higher dimensional case.} This means that we have
the equivalence of the two bulk/boundary actions given by the
respective gauge transformations $c \mapsto c - d\alpha$ and
$\pi^\sharp \mapsto
\pi^\sharp\circ(1+\alpha_\sharp\circ\pi^\sharp)^{-1}$.

\subsubsection*{Poisson sigma model}

On shell, using the equations
of motion for $\boldsymbol{F}$'s, we obtain the closed twisted
Poisson sigma model \cite{Park}, \cite{KlimcikStrobl}
$$S_{(\pi,c}) =\int_{\partial X} ({\boldsymbol{\chi}}_i {D\boldsymbol{X}}^i+
\frac{1}{2}\boldsymbol{\pi}^{ij}{\boldsymbol{\chi}}_i
{\boldsymbol{\chi}}_j) + \frac{1}{6}\int_X
\boldsymbol{c}_{ijk}D\boldsymbol{X}^{i}D\boldsymbol{X}^{j}D\boldsymbol{X}^{k}\,.$$

\begin{remark}\label{open}
In order to obtain the open (twisted) Poisson sigma model,
in \cite{HofmanPark} it is proposed to include boundaries with
corners and allow for different boundary conditions on various
regions of the boundary. If, for example, the boundary $\partial X$
is divided in two regions, on the first one takes the same boundary
conditions as before and on the second one restricts the superfields
to only $V$. The interface of the two regions can then be viewed as
the boundary of the first region. On the interface live only
superfields corresponding to functions on $V$, as on the boundary of
the Poisson sigma model.
\end{remark}

\section{Open topological $p$-branes and Nambu-Poisson sigma models}

Let, again, $V$ be a smooth manifold. We put
$M=T^*[p]((\wedge^{p-1}T)[p-1](T[1]V))$,
with $p\geq2$ an integer. The
canonical symplectic structure $\omega$ on the target $M$ is of
degree $p$. Hence, we take a $p+1$ dimensional $\Sigma$. We will
denote the degree 0 local coordinates on $V$ as $X^i$, the induced
degree 1 fibre coordinates on $T[1]V\to V$ by $\psi^i$, the induced
degree ${p-2}$  and degree ${p-1}$ fibre coordinates on
$(\bigwedge^{p-1}T)[p-1](T[1]V)\to T[1]V$ as
$H^{I}:=H^{i_1\ldots i_{p-1}}$, with $i_1<\ldots< i_{p-1}$ and
$\eta^I:=\eta^{i_1\ldots i_{p-1}}$, $i_1<\ldots< i_{p-1}$, respectively.
Further, the dual fibre coordinates on
$T^*[p]((\bigwedge^{p-1}T)[p-1](T[1]V))\to (\bigwedge^{p-1}T)[p-1](T[1]V)$ of the
respective degrees $p-1$, $p$, 2 and 1 will be denoted by $\chi_i$,
$F_i$, $G_I:=G_{i_1\ldots i_{p-1}}$, $i_1<\ldots< i_{p-1}$, and
$A_I:=A_{i_1\ldots i_{p-1}}$, $i_1<\ldots< i_{p-1}$. The canonical
symplectic form in these coordinates is
$$\omega= dF_i\wedge dX^i +
d\psi^i \wedge d\chi_i+  dG_I\wedge dH^I + d\eta^I\wedge dA_I\,.$$
The potential one-form $\vartheta$ can be taken as $\vartheta = F_idX^i
+ \psi^id\chi_i + G_IdH^I  + \eta^IdA_I$. Its zero locus $L$ is
given by $F_i=0$, $\psi^i=0$, $G_I=0$ and $\eta_I=0$. We choose the
submanifold $L'\subset L$ of $L$ by letting $H^I=0$ and requiring
that all the products $\chi_iA_{i_1\ldots i_{p-1}}$ are totally
antisymmetric in all their indices.

We may use the superfields $\boldsymbol{X}^i$, $\boldsymbol{F}_i$,
$\boldsymbol{\psi}^i$, $\boldsymbol{\chi}_i$, $\boldsymbol{H}^I$,
$\boldsymbol{G}_I$, $\boldsymbol{\eta}_I$ and $\boldsymbol{A}_I$ to
write the BV bracket in the form
$$\int_X (\partial_{{\boldsymbol{X}}^i}\wedge
\partial_{{\boldsymbol{F}}_i}
+\partial_{{\boldsymbol{\chi}}_i}\wedge
\partial_{{\boldsymbol{\psi}}^i}+ \partial_{{\boldsymbol{H}}^I}\wedge
\partial_{{\boldsymbol{G}}_I}+ \partial_{{\boldsymbol{A}}_I}\wedge\partial_{{\boldsymbol{\eta}}^I}) \,.
 $$
For the $S_0$ part of the BV action we have
$$S_0=\int_X ({\boldsymbol{F}}_i{D\boldsymbol{X}}^i +
{\boldsymbol{\psi}}^i{D\boldsymbol{\chi}}_i +
\boldsymbol{G}_ID{\boldsymbol{H}}^I +
{\boldsymbol{\eta}}^I{D\boldsymbol{A}}_I)\,.$$
Furthermore, we can now add  the Hamiltonian function $-\psi^iF_i$
corresponding to the de Rham differential on $V$  and further terms
not depending on a background, so that the corresponding degree
$(p+1)$ function $$\gamma_0=-\psi^iF_i + \frac{1}{(p-1)!}G_{i_1\ldots
i_{p-1}}(\eta^{i_1\ldots
i_{p-1}}-\psi^{i_1}\ldots\psi^{i_{p-1}})\,,$$ is still a
solution to the classical master equation on $M=T^*[p]( (\bigwedge^{p-1}T)[p-1](T[1]V))$
(cf.\ the remark at the end of Section \ref{DorfmanRemark}), satisfying $\gamma_0|_L=0$.
Hence, we have the following BV action
\begin{align*}
S_0+ \Gamma_0 &=\int_X ({\boldsymbol{F}}_i{D\boldsymbol{X}}^i +
{\boldsymbol{\psi}}^i{D\boldsymbol{\chi}}_i +
\boldsymbol{G}_ID{\boldsymbol{H}}^I +
{\boldsymbol{\eta}}^I{D\boldsymbol{A}}_I) \\&+ \int_X
(-\boldsymbol{\psi}^i \boldsymbol{F}_i +
\frac{1}{(p-1)!}\boldsymbol{G}_{i_1\ldots
i_{p-1}}(\boldsymbol{\eta}^{i_1\ldots i_{p-1}}-
\boldsymbol{\psi}^{i_1}\ldots\boldsymbol{\psi}^{i_{p-1}})) \,.
\end{align*}

\subsubsection*{Bulk interaction}\label{bulk}

Let $c$ be a $(p+1)$-form on $V$. In
local coordinates, \\ \noindent $c=\frac{1}{(p+1)!}c_{i_1\ldots
i_{p+1}}dX^{i_1}\wedge \ldots \wedge dX^{i_{p+1}}$. We associate
with it the degree $p+1$ function $C=\frac{1}{(p+1)!}{c}_{i_1\ldots
i_{p+1}}\psi^{i_1}\ldots \psi^{i_{p+1}}$ on $M$ and the corresponding
degree $0$ function $\Gamma_1$ on $\mathcal{M}$. By
construction,
$$\boldsymbol{\{}S_0 + \Gamma_0 + \Gamma_1, S_0 + \Gamma_0 +
\Gamma_1 \boldsymbol{\}}
=2  \boldsymbol{\{}\Gamma_0, \Gamma_1 \boldsymbol{\}}\,.$$ Obviously, the sum
\begin{align}
S_0 + \Gamma_0 + \Gamma_1 &= \int_X
({\boldsymbol{F}}_i{D\boldsymbol{X}}^i +
{\boldsymbol{\psi}}^i{D\boldsymbol{\chi}}_i +\boldsymbol{G}_ID\boldsymbol{H}^I +
{\boldsymbol{\eta}}^I{D\boldsymbol{A}}_I)\nonumber\\
&+ \int_X(-\boldsymbol{\psi}^i \boldsymbol{F}_i +
\frac{1}{(p-1)!}\boldsymbol{G}_{i_1\ldots
i_{p-1}}(\boldsymbol{\eta}^{i_1\ldots i_{p-1}}-
\boldsymbol{\psi}^{i_1}\ldots\boldsymbol{\psi}^{i_{p-1}}))\nonumber\\&+
\frac{1}{(p+1)!}\int_X\boldsymbol{c}_{i_1\ldots
i_{p+1}}\boldsymbol{\psi}^{i_1}\ldots
\boldsymbol{\psi}^{i_{p+1}}\nonumber
\end{align} is a solution to the master equation iff the
$(p+1)$-form $c$ is closed. Let us also note that the canonical
transformation on $M$, generated by the degree $p$ function
${\alpha}$, $\alpha_{|L'}=0$, $\{\alpha,\alpha\}=0$, where
$\alpha=\frac{1}{p!}({b}_{i_1\ldots
i_{p}}\psi^{i_1}{\eta}^{i_2\ldots i_{p}} -
\partial_{i_1}{b}_{i_2\ldots
i_{p+1}}\psi^{i_1}\psi^{i_2} {H}^{i_3\ldots i_{p+1}})$, with $b$
being a $p$-form on $V$, amounts into the gauge transformation
$c\mapsto c - db$. Such a canonical transformation preserves
$L$ as well as $L'$.

\subsubsection*{Boundary interaction}

Again, we can allow for
a nonempty boundary $\partial \Sigma\neq\emptyset$ of $\Sigma$ and
try to add a boundary interaction. The boundary conditions can be
chosen so that the superfields restrict on the boundary to maps to
the submanifold $L'\subset L$ of the zero locus $L$ of $\vartheta$.
This means that for the superfields $\boldsymbol{F}_i$,
$\boldsymbol{\psi}^i$, $\boldsymbol{G}_I$ and $\boldsymbol{\eta}^I$
we take Dirichlet boundary conditions, as well as for
$\boldsymbol{H}^I$ and
$(\boldsymbol{\chi}_{i_1}\boldsymbol{A}_{i_2\ldots
i_{p}}+\boldsymbol{\chi}_{i_k}\boldsymbol{A}_{i_2\ldots
i_{k-1}i_1i_{k+1} \ldots i_{p}})$.

We can consider a boundary term associated to a $p$-vector field $\pi$
on $V$. This can be done considering the canonical transformation on
$M$, now generated by the degree $p$ function
$-\frac{1}{(p-1)!}{\pi}^{i_1 i_2\ldots i_{p}}{\chi}_{i_1}A_{i_2\ldots
i_{p}}=-\frac{1}{(p-1)!}{\pi}^{i_1 i_2\ldots i_p}A_{i_1\ldots
i_{p-1}}{\chi}_{i_p}$. We know that the result is equivalent to the
boundary/bulk BV action
\begin{align}\label{SCPI}
S_{c,\pi}&=S_0 + \Gamma_0 + \Gamma_1 = \int_X
({\boldsymbol{F}}_i{D\boldsymbol{X}}^i +
{\boldsymbol{\psi}}^i{D\boldsymbol{\chi}}_i
+{\boldsymbol{G}}_I{D\boldsymbol{H}}^I +
{\boldsymbol{\eta}}^I{D\boldsymbol{A}}_I)\nonumber\\ &+
\int_X(-\boldsymbol{\psi}^i \boldsymbol{F}_i +
\frac{1}{(p-1)!}\boldsymbol{G}_{i_1\ldots
i_{p-1}}(\boldsymbol{\eta}^{i_1\ldots i_{p-1}}-
\boldsymbol{\psi}^{i_1}\ldots\boldsymbol{\psi}^{i_{p-1}}))
\nonumber\\&+ \frac{1}{(p+1)!}\int_X\boldsymbol{c}_{i_1\ldots
i_{p+1}}\boldsymbol{\psi}^{i_1}\ldots \boldsymbol{\psi}^{i_{p+1}}
+\int_{\partial X}\frac{1}{(p-1)!}\boldsymbol{\pi}^{i_1 i_2\ldots
i_p}\boldsymbol{A}_{i_1\ldots i_{p-1}}{\boldsymbol{\chi}}_{i_p} \,.
\end{align}
For $\gamma= -\psi^i F_i + \frac{1}{(p-1)!}G_{i_1\ldots i_{p-1}}(\eta^{i_1\ldots
i_{p-1}}- \psi^{i_1}\ldots\psi^{i_{p-1}}) +
\frac{1}{(p+1)!}{c}_{i_1\ldots i_{p+1}}\psi^{i_1}\ldots
\psi^{i_{p+1} }$, the condition $\{\gamma, \gamma\}=0$ gives, as
before, $dc=0$. Regarding the condition $\gamma|_{L'_\pi}=0$, we
notice that the Lagrangian submanifold ${L_\pi}$ is given by
equations
\begin{equation}\label{cond1}\psi^i= \{{\pi}^{Jj}A_J{\chi}_j, \psi^i\}=
\pi^{Ji}A_J\,, \end{equation}
\begin{equation}\label{cond2}\eta^I= \{{\pi}^{Ji}A_J{\chi}_i,
\eta^I\}= \pi^{jI}\chi_j\,, \end{equation}
\begin{equation}\label{cond3}G_I=0\,, \end{equation} and
\begin{equation}\label{cond4}F_i=\{{\pi}^{Jj}A_J{\chi}_j, F_i\}=
\partial_i\pi^{Kj}A_K\chi_j\,.\end{equation} Also, the conditions
$H^I=0$ and the products $\chi_{i_1}A_{i_2\ldots i_p}$ being totally
antisymmetric in its indices are not affected by this canonical
transformation. Hence, from $\gamma|_{L'_\pi}=0$, it follows that
\begin{equation}
-\pi^{Ji}\partial_i(\pi^{Kj}) A_J A_K\chi_j +
\frac{1}{(p+1)!}{c}_{i_1\ldots i_{p+1}}\pi^{J_1i_1}\ldots
\pi^{J_{p+1}i_{p+1}}A_{J_1}\ldots A_{J_{p+1}}=0\,.\label{boundary}
\end{equation}
If we now assume a (locally) decomposable $\pi$, the second term
will vanish automatically. Taking into account that
$\chi_{i_1}A_{i_2 \ldots i_{p}}$ are totally antisymmetric in all
their indices, for a (locally) decomposable $\pi$, this equation
gives the differential condition (\ref{alg2}).

Before we summarize the above discussion, let us note that the
decomposability assumption  of $\pi$ is quite natural as the
following remarks show.
\begin{remark}\label{antisymm}
With a decomposable $\pi$, on $L'_\pi\subset L_\pi$ also the
products $\psi^{i_1}\eta^{i_2\dots i_{p-1}}$ are antisymmetric in
all their indices. Using the decomposability of $\pi$ in the form of Lemma
(\ref{decompLemma}) and the antisymmetry of the products $\chi_{i_1}A_{i_2
\ldots i_{p}}$ it is easy to see that $p\psi^{i_1}\eta^{i_2\dots
i_{p-1}}= \pi^{j_1\ldots j_p}A_{j_1\ldots
j_{p-1}}\chi_{j_p}\pi^{i_1\ldots i_p}$, from where the claim
follows. \end{remark}

\begin{remark} \label{necessary}
Also, the decomposability of  $\pi$ -- at a point of $V$ where $\pi$
is nonzero -- is a necessary condition for the second term to vanish
independently of $c$. From the condition
$\pi^{J_1i_1}\ldots\pi^{J_{p+1}i_{p+1}}A_{J_1}\ldots A_{J_{p+1}}=0$
it follows that (locally) the rank of the map
$\pi^\sharp:\Omega^{p-1}(V)\to \mathfrak X(V)$ has to be smaller
than $(p+1)$. On the other hand, since $\pi$ is of order $p$ the
rank of this map has to be at least $p$. Hence, the rank of
$\pi^\sharp:\Omega^{p-1}(V)\to \mathfrak X(V)$ is $p$. This is
equivalent to the decomposability of $\pi$.
\end{remark}

Now we can summarize the above discussion.
\begin{theorem} Let $L_\pi$ be the Lagrangian submanifold given by
$F_i=\psi^i=\eta^I=G_I=0$ and let $L'_\pi\subset L_\pi$ be its
submanifold given by the conditions $H^I=0$ and  $\chi_{i_1}A_{i_2
\ldots i_{p}}+ \chi_{i_k}A_{i_2 \ldots i_{k-1}i_1 i_{k+1}\ldots
i_{p}}=0$. Also, let the superfields restrict at the boundary to
maps into a submanifold $L'_\pi\subset L_\pi$. Let, further, $c$ be
a closed $p+1$-form and $\pi$ a Nambu-Poisson tensor of order $p$,
for $p\geq 2$. Then the bulk/boundary action $S_{c,\pi}$ of Eqn.~(\ref{SCPI})
is a BV action.\footnote{For $p=2$, we can also assume
a not necessarily decomposable $\pi$ and  choose consistently the
submanifold $L'_\pi$ of $L_\pi$ given by additional conditions
$A_i=\chi_i$. Then the condition (\ref{boundary}) means that $\pi$
is a Poisson structure twisted by $c$.}
\end{theorem}

For the converse statement, see Remark \ref{necessary} above.
We finish this section with the following remark.
\begin{remark}
Let us note that the equation (\ref{cond1}) gives the graph of
$\pi^\sharp: \Omega^{p-1}(V)\to \mathfrak X(V)$ in $\mathfrak
X(V)\oplus \Omega^{p-1}(V)$ and the equation (\ref{cond2}) the graph
of the map dual to $\pi^\sharp$. Let us also note that the condition
(\ref{cond2}) was not used at all in order to derive the
integrability condition (\ref{boundary}). In \cite{Hagiwara} and
\cite{Wade} higher Dirac structures (Nambu-Dirac structures)
on $\mathfrak X(V)\oplus \Omega^{p-1}(V)$, or more generally on
$A\oplus \wedge^{p-1}A^*$, $A$ being a Lie algebroid, were defined.
The graph of the map $\pi^\sharp$ corresponding to a $p$-tensor
$\pi$ is an example of a higher Dirac structure iff $\pi$ is a
Nambu-Poisson tensor. It would be interesting to further explore,
similarly to \cite{HofmanPark}, the relation between boundary
conditions and higher Dirac structures.
\end{remark}

\subsubsection*{Gauge transformations}

Recall Corollary \ref{MainGaugeTr}.
Consider the degree $p$ function ${\alpha}$, $\alpha_{|L'}=0$,
$\{\alpha,\alpha\}=0$, where $\alpha=\frac{1}{p!}({b}_{i_1\ldots
i_{p}}\psi^{i_1}{\eta}^{i_2\ldots i_{p}} -
\partial_{i_1}{b}_{i_2\ldots
i_{p+1}}\psi^{i_1}\psi^{i_2} {H}^{i_3\ldots i_{p+1}})$, with $b$
being a $p$-form on $V$. Let us recall that the canonical
transformation generated by ${\alpha}$, results in the gauge
transformation $c\mapsto c - db$. Such a canonical transformation
preserves both $L$ as well as $L'$.

Also, consider the degree $p$  function $\beta$,
$\{\beta,\beta\}=0$, given as \\
\noindent
$\beta= \frac{1}{(p-1)!}{\pi}^{i_1 i_2\ldots
i_p}{\chi}_{i_1}A_{i_2\ldots i_{p}}$, with $\pi$ being a
Nambu-Poisson tensor. We are looking for the
solution to the factorization problem $e^{\delta_\alpha}
e^{\delta_\beta}= e^{\delta_{\beta'}}e^{\delta_{\alpha'}}$ (cf. Corollary \ref{MainGaugeTr}). In
order to solve it, we can follow the strategy of \cite{HofmanPark} in the case
of  the topological open membrane. For this we replace $\alpha$ by
$t\alpha$. Then $\alpha'$ and $\beta'$ will depend on $t$ and
therefore will be denoted as $\alpha_t'$ and $\beta_t'$,
respectively. Obviously, $\beta_0'= \beta$. Inspired by Corollary \ref{EquivCoro}, we
take $\beta_t'$ again of the form $\beta'=
\frac{1}{(p-1)!}{\pi'_t}^{i_1 i_2\ldots i_p}{\chi}_{i_1}A_{i_2\ldots
i_{p}}$, with ${\pi'_t}$ an antisymmetric $p$-tensor field. We
have
$$\frac{d}{dt}(e^{-\delta_{\beta_t'}}e^{t\delta_\alpha}
e^{\delta_\beta})=e^{-\delta_{\beta_t'}}\delta_\alpha
e^{t\delta_\alpha}
e^{\delta_\beta}-\delta_{\dot\beta_t'}e^{-\delta_{\beta_t'}}e^{t\delta_\alpha}
e^{\delta_\beta}= \delta_{\epsilon_t}
e^{-\delta_{\beta_t'}}e^{t\delta_\alpha} e^{\delta_\beta}\,,
$$
with
\begin{align*}
{\epsilon_t} & =e^{-\delta_{\beta_t'}}\alpha-\dot\beta_t'
 \\ & =
\frac{1}{p!}({b}_{i_1\ldots i_{p}}\psi_t^{i_1}{\eta_t}^{i_2\ldots
i_{p}} -
\partial_{i_1}{b}_{i_2\ldots
i_{p+1}}\psi_t^{i_1}\psi_t^{i_2} {H}^{i_3\ldots
i_{p+1}})-\frac{1}{(p-1)!}{\dot{\pi'}_t}^{i_1 i_2\ldots
i_p}A_{i_1\ldots i_{p-1}}{\chi}_{i_p} \,,
\end{align*}
where
$$\psi_t^i= \psi^i -
{\pi'}_t^{Ji}A_J\,,$$ and
$$\eta_t^I=\eta^I + (-1)^p
{\pi'}_t^{Ij}\chi_j\,.$$

Since we ask $\epsilon_t$ to vanish on $L'$, the terms proportional
to the products $\chi A$ have to be zero. The resulting differential
equation for $\pi'_t{}^\sharp$, with the initial condition $\pi_0^\sharp=\pi^\sharp$ has
the solution\footnote{See Remark \ref{diffequation}.}
$${\pi'}^\sharp_t=\pi^\sharp\circ(1 + b_\sharp\circ\pi^\sharp)^{-1}=
(1 + (-1)^{p-1}b(\pi))^{-1}\pi^\sharp\,.$$ Hence, $\pi'=\pi_1'$ is a
Nambu-Poisson tensor, gauge equivalent to $\pi$. Fortunately, we do
not need the the explicit form of $\alpha'$, which is the solution
to $\frac{d}{dt}e^{\delta_{\alpha_t'}}=
\delta_{\epsilon_t}e^{\delta_{\alpha_t'}}$, $\alpha'_0=0$. It is
enough to know that $\alpha'|_{L'}=0$ which guaranteed by
$\epsilon_t$ vanishing on $L'$.

Let us finish this subsection by recalling that in the case of an exact $b$ the gauge
transformation between $\pi$ and $\pi'$
can be identified as the Seiberg-Witten map, see Proposition \ref{SW} ff.

\subsubsection*{Nambu-Poisson sigma model}

On shell, using the equations of motion for $\boldsymbol{F}$'s and
$\boldsymbol{G}$'s, we obtain the closed (twisted) Nambu-Poisson
sigma model, cf. also \cite{JS}
\begin{align}S_{(\pi,c)} &=\int_{\partial X} {\boldsymbol{\chi}}_i {D\boldsymbol{X}}^i
\nonumber\\ &+ \int_{\partial
X}(\frac{1}{(p-1)!}\boldsymbol{A}_{i_1\ldots i_{p-1}}
D\boldsymbol{X}^{i_1}\ldots D\boldsymbol{X}^{i_{p-1}}+
\frac{1}{(p-1)!}{\pi}^{i_1 i_2\ldots i_p}\boldsymbol{A}_{i_1\ldots
i_{p-1}}{\boldsymbol{\chi}}_{i_p})\nonumber\\ &+
\int_X\frac{1}{(p+1)!}{c}_{i_1\ldots
i_{p+1}}{D\boldsymbol{X}}^{i_1}\ldots {D\boldsymbol{X}}^{i{p+1}} \,.
\end{align}

\begin{remark}
Let us note that from equations of motion for $\boldsymbol{\chi}$'s
and $\boldsymbol{A}$'s
$$D\boldsymbol{X}^i= \frac{1}{(p-1)!} \boldsymbol{\pi}^{iJ}\boldsymbol{A}_J \,,$$
and
$$D\boldsymbol{X}^{i_1}\ldots D\boldsymbol{X}^{i_{p-1}}
=\boldsymbol{\pi}^{i_1\ldots i_{p-1}i_p}\boldsymbol{\chi}_{i_p}\,.$$
it follows that the products
$$ \pi^{iJ}A_J \pi^{i_1\ldots i_{p-1}i_p}\chi_{i_p}$$ must be antisymmetric
in indices $i,i_1\ldots, i_{p-1}$, which is consistent due to the
decomposability of $\pi$ and the antisymmetry of the products
$\chi_i A_{i_1\ldots i_{p-1}}$ (cf. Remark \ref{antisymm}).
\end{remark}

\begin{remark}
In order to obtain the open Nambu-Poisson sigma
model, we can follow the idea of \cite{HofmanPark}, cf. Remark
\ref{open}, and include boundaries with corners and allow for
different boundary conditions on various regions of the boundary.
If, for example, the boundary $\partial X$ is divided in two
regions, on the first one takes the same boundary conditions as
before and on the second one restricts the superfields to only $V$.
The interface of the two regions can then be viewed as the boundary
of the first region. On the interface live only superfields
corresponding to functions on $V$.
\end{remark}


\subsection*{Acknowledgments}
B.J. would like to thank Ivo Sachs and Peter Schupp for discussions
and IHES, LMU and MPIM for hospitality. The research of P.B. was
supported under the Australian Research Council's Discovery Projects
funding scheme (project numbers DP0878184 and DP110100072) and the
research of B.J. under Institutional grant MSM 0021620839.



\end{document}